\documentclass[onecolumn]{emulateapj}
\usepackage{natbib}
\usepackage{amsmath}
\usepackage{epstopdf}

\slugcomment{Submitted to The Astrophysical Journal, 10/30/14}

\shorttitle{Planet masses with TTV}
\shortauthors{Deck, Agol}

\begin{document}


\title{Measurement of planet masses with transit timing variations due to synodic ``chopping'' effects}


\author{Katherine M. Deck}
\affil{Division of Geological and Planetary Sciences, California Institute of Technology, Pasadena, CA 91101}
\email{kdeck@caltech.edu}
\and
\author{Eric Agol}
\affil{Department of Astronomy, University of Washington, Box 351580,
    Seattle, WA 98195}
\email{agol@astro.washington.edu}



\begin{abstract}
Gravitational interactions between planets in transiting exoplanetary systems lead to 
variations in the times of transit that are diagnostic of the planetary masses and the dynamical
state of the system.  Here we show that synodic
``chopping" contributions to these transit timing variations (TTVs)
can be used to uniquely measure the masses of planets without full dynamical analyses involving direct integration of the equations of motion. We present simple analytic formulae
for the chopping signal, which are valid (generally $<10\%$ error) for modest eccentricities $e \la 0.1$.  
Importantly, these formulae primarily depend on the mass of the perturbing planet, and therefore the chopping signal can be used to break the mass/free-eccentricity degeneracy which can appear for systems near first order mean motion resonances. Using a harmonic analysis, we apply these TTV formulae to a number of {\it Kepler} systems which had been previously analyzed with full dynamical analyses. We show that when chopping is measured, the masses of both planets can be determined uniquely, in agreement with previous results, but {\it without}  the need for
numerical orbit integrations.  This demonstrates how mass measurements from TTVs may primarily arise from an observable chopping signal. The formula for chopping 
can also be used to predict the number of transits and timing precision required for future observations, such as those made by TESS or PLATO, in order to infer planetary masses through analysis of TTVs.
\end{abstract}





\section{Introduction}

In a multi-planet system, mutual gravitational interactions between planets
lead to deviations from Keplerian orbits. In particular, the instantaneous orbital periods are no longer constant, which in turn implies that transiting planets in multi-planet systems will not transit at a fixed, constant rate. The detection of these changes in the transit rate,  or `transit-timing
variations' (TTVs), was initially recognized as a way to infer the presence of non-transiting planets in systems with at least one other transiting planet  \citep{Schneider2003,Agol2005,Holman2005,Miralda2002}.
TTVs have since been used to confirm that a transit light curve signal is due to a planetary transit \citep[e.g.][]{Kepler9}, to 
constrain planetary orbital elements and measure planetary masses using photometry alone \citep[e.g.][]
{CarterAgol2012}, and to detect and characterize non-transiting planets \citep[e.g.][]{Ballard,KOI872}.

TTV data are most commonly analyzed through inversion, a process through which observed transit times are fit using a model of gravitationally interacting planets in order to determine the system parameters, including planetary masses relative to the mass of the star, as well as orbital elements.\footnote{The relative frequency of TTVs in different types of systems has also been analyzed \citep{Xie2014}.} Transiting exoplanets generally have well constrained radii, so measurements of their masses yields information regarding densities,
bulk compositions and gravities. This in turn can be used to identify promising targets for atmospheric characterization and to constrain
planetary formation and dynamical evolution \citep[e.g.][]{HansenMurray2013}. Of the orbital elements, constraints on the planetary eccentricities in particular are necessary to understand the importance of interaction with the protoplanetary disk, interactions with remnant planetesimals, and tidal dissipation \citep[e.g.][]{ResonantRepulsion,HansenMurray2013,Batygin2013,Hansen2014,Mahajan2014}.

However, the TTV inversion problem is often complicated by strong nonlinear correlations between parameters in a large dimensional space, and as a result precise planetary mass and orbit measurements can be difficult to make. Many of the {\it Kepler} multi-planet systems with partially characterized planetary orbits and masses are those near first order mean motion resonances, a configuration in which the period ratio of two planets is close to $P_1/P_2
\approx (j_R-1)/j_R$, where $j_R$ is an integer greater than unity,  $P_1$ is the period of the 
inner planet, and $P_2$ is the period of the outer planet. Indeed, for nearly circular orbits and given planet-to-star mass ratios, TTVs are largest in amplitude near first-order mean motion resonances \cite[e.g.][]{Agol2005,Holman2005}.  If the planets are near to, but not in resonance, then 
the planets show sinusoidal variations with a period equal to the `super-period',
\begin{equation}
P^j = {1 \over \vert j_R/P_2 - (j_R-1)/P_1\vert}
\end{equation}
\citep{Agol2005,Lithwick2012}.  However, the
amplitude of this TTV signal depends on both the mass of the perturbing planet
and the eccentricity vectors of both planets \citep{Lithwick2012}.  This degeneracy can 
be broken statistically with analyses of a large number of planetary systems
\citep{HaddenLithwick} or for systems with very precisely measured transit times; however in practice it inhibits the measurement of the 
masses and limits our knowledge of the eccentricities of individual planetary systems.

In spite of this mass-eccentricity degeneracy (and others\footnote{Degeneracies have also been identified between eccentricity vector components and between planetary mass and mutual inclination \citep{Lithwick2012,CarterAgol2012,Kep56}.}), it has been possible in some cases to precisely measure the masses of planets using TTVs \citep[e.g.][]{CarterAgol2012,KOI142,Masuda,Kep9revisit, Nesvorny2014a}. The successful mass measurements in these systems is due to the fact that an additional, independent periodic component of the TTVs, with a timescale other than the super-period, was resolved. Other components of TTVs have amplitudes that depend on the orbital parameters and masses in different ways, and so the measurement of secondary components leads to additional, independent constraints on orbital parameters. In particular, the so-called short-timescale\footnote{The short-timescale variations occur on harmonics of the synodic timescale, $(P_1^{-1}-P_2^{-1})^{-1}$, but can sometimes appear to vary on a longer timescale due to aliasing.} ``chopping" TTV associated with the planetary synodic timescale has been identified as an important feature for unique characterization of systems \citep{Kepler9,KOI142}. More recently, \citet{Nesvorny2014} studied this chopping TTV to clarify how, despite degeneracies between parameters, the TTV method can be used to measure planetary masses in the case of low-eccentricity orbits. 

In Section \ref{sec:harmonic}, we begin by describing a harmonic approach to analyzing TTVs.  We review the work of \citet{Lithwick2012} in Section \ref{sec:firstorder}, and discuss the TTV signal for a system near a first order mean motion resonance (referred to hereafter as the ``Lithwick et al. formula"). We then introduce the conjunction effect and give analytic formulae for the chopping signal in Section \ref{sec:chopping}; these were derived first in \citet{Agol2005} and more recently, using a similar approach, in \citet{Nesvorny2014}. We show how the chopping formula encompasses near-resonant effects, and more generally consider the range of validity of the synodic TTV formulae. We then place the synodic TTV expression in context with that of Lithwick et al. and discuss the regimes in which each should be used. In Section \ref{subsec:predict}, we address the more general problem of using these formulae to predict, given the timing precision on the transits, how many observations are required to infer the masses of a particular system. This will be important in the planning of follow-up observations of partially characterized systems and for estimating the timing precision required for future surveys to obtain a measurement of planetary masses through chopping.

In Section \ref{sec:applications}, we apply the synodic TTV formula to {\it Kepler} data, and use it to infer planetary masses for systems both near and far from mean motion resonance. The synodic formulae can be used alone to determine planetary masses, or in combination with the Lithwick et al. formula for systems near first order mean motion resonances, in  which case the mass-free eccentricity degeneracy can be broken and a constraint on the free eccentricities can be determined as well. We present our conclusions in Section \ref{sec:conclusion}. In the Appendix we give an alternate derivation of the synodic TTV formulae based on Hamiltonian perturbation theory, and we discuss the convergence of the series for the synodic chopping signal.

\section{Transit aliasing and harmonic analysis of TTVs}\label{sec:harmonic}
The TTVs of a planet can be written as a combination of  periodic components with frequencies that are integer combinations of
the two interacting planets' orbital frequencies, $n_{pq} = p n_1 - q n_2$, where
$(p,q)$ are integers, and $n_{1,2}=2\pi/P_{1,2}$ are the mean motions
of the two planets \citep{NesvornyMorbidelli2008, N3,NB10}.  More explicitly,
transit-timing variations for a two planet system can be expanded as:
\begin{eqnarray}
\delta t_{1,k} &=& P_1 \frac{m_2}{M_\star} \sum_{p,q} \bigg[a_{1,p,q} \cos{[(p n_1 - q n_2) t_{1,k}]} + b_{1,p,q} \sin{[(p n_1 - q n_2) t_{1,k}]}\bigg], \cr
\delta t_{2,k} &=& P_2  \frac{m_1}{M_\star} \sum_{p,q} \bigg[a_{2,p,q} \cos{[(p n_1 - q n_2) t_{2,k}]} + b_{2,p,q} \sin{[(p n_1 - q n_2) t_{2,k}]}\bigg],
\end{eqnarray}
where $k$ denotes the transit number, $a_{i,p,q},b_{i,p,q}$ ($i=1,2$) are coefficients which are functions of
the orbital elements of the planets (except the mean longitudes $\lambda_i$), and therefore vary on timescales long compared to the orbital period; $m_1,m_2$ are the masses of the two planets; $M_\star$ is the mass of the star; and $t_{i,k}$ is the $k$th transit time of the $i$th planet. We  assume that the observation baseline is short compared to the secular timescales (which are typically ``long" since they are proportional to $P_i M_\star/m_{planet}$) so that treating the coefficients $a_{i,p,q},b_{i,p,q}$ as constant is justified. We will discuss this further in Section \ref{sec:EccEff}.   Note that the transit
timing variations scale in proportion to the orbital period of each
transiting planet and in proportion to the mass ratio of the perturbing
planet. (These equations and scaling relations do not apply in mean-motion
resonance \citep{Agol2005}.)

The
transit times, $t_{i,k}$ are converted to transiting timing variations after removing a mean
ephemeris, $\delta t_{i,k} = t_{i,k} - t_{i,0} - P_i  k$, where $k$ is an integer \citep{Agol2005}. Since transit timing variations are typically much smaller than 
the planetary orbital periods,  $\vert \delta t_{i,k} \vert \ll P_i$, 
the planets' transits are (nearly) sampled on their orbital frequencies, so
\begin{eqnarray}
\delta t_{1,k} &\approx& P_1  \frac{m_2}{M_\star} \sum_{p,q} \bigg[a_{1,p,q} \cos{[(p n_1 - q n_2) (t_{1,0}+P_1 k)]} + b_{1,p,q} \sin{[(p n_1 - q n_2) (t_{1,0}+P_1 k)]}\bigg], \cr
\delta t_{2,k} &\approx& P_2  \frac{m_1}{M_\star} \sum_{p,q} \bigg[a_{2,p,q} \cos{[(p n_1 - q n_2) (t_{2,0}+P_2 k)]} + b_{2,p,q} \sin{[(p n_1 - q n_2) (t_{2,0}+P_2 k)]}\bigg],
\end{eqnarray}
where we have dropped terms of order $(m_{planet}/M_\star)^2$. 
However, since $n_i P_i = 2\pi$, these equations can be rewritten as:
\begin{eqnarray}
\delta t_{1,k} &\approx& P_1  \frac{m_2}{M_\star} \sum_{q} \bigg[a^\prime_{1,q} \cos{2\pi q (P_1/P_2) k} + b^\prime_{1,q} \sin{2\pi q (P_1/P_2) k)}\bigg], \cr
\delta t_{2,k} &\approx& P_2  \frac{m_1}{M_\star} \sum_{p} \bigg[a^\prime_{2,p} \cos{2\pi p (P_2/P_1) k} + b^\prime_{2,p} \sin{2\pi p (P_2/P_1) k)}\bigg],
\end{eqnarray}
where, e.g. for the inner planet, the sum over $p$ is now implicit: for a specific value of $q$, the coefficient for each possible value of $p$ has
been absorbed into the new coefficients $a^\prime, b^\prime$ (for the outer planet, the sum over $q$ is now implicit).  Therefore, each coefficient
now contains variations due to multiple linearly independent frequencies of the form $(p n_1 - q n_2)$. But because of the sampling of the TTV on the transiting planet's orbital period,  frequencies that differ by integer multiples
of the orbital frequency of the transiting planet are indistinguishable in transit timing variations. Therefore short-period TTVs (arising from short-timescale periodic terms in the variation of the orbital elements) can contribute to the same harmonic of the perturbing planet as resonant TTVs, those arising from any near-resonant configuration. For example, in a system near the 2:1 resonance, the fast frequency $2(n_1-n_2)$ will, due to the discrete sampling at every transit of the inner planet, be aliased such that it appears as a sinusoid with the super-period, i.e. the same timescale as the resonant frequency $2n_2-n_1$ (akin to the stroboscopic effect in which a spinning car wheel appears to be rotating at a slower rate due to the sampling rate).

The inner planet will have apparent TTV frequencies that are integer 
multiples of the perturbing planet's orbital frequency, $ q n_2$, while the outer planet
TTVs will show frequencies that are integer multiples of the inner
planet's frequency, $p n_1$.  If the perturbing planet does not transit the host star, this aliasing can mask the true period of the perturbing planet since we can add integer multiples of the transiting planet's orbital frequency without affecting the TTV.  If both 
planets transit the star, though, then both periods and orbital phases are 
known.   Consequently, TTVs can be fit using a harmonic
analysis of the perturbing planet's orbital frequency if that planet has been
identified.  We utilize this harmonic analysis below to identify components
of the TTV that are caused by the conjunctions 
of the planets \citep{Nesvorny2014}, and thus depends on the difference between their 
mean longitudes, the synodic angle $\psi = \lambda_1 -\lambda_2 = (n_1-n_2) t + \psi(t=0)$.
But first we turn to a review of TTVs due to first-order resonant terms.

\section{First-Order Resonant TTV signal}\label{sec:firstorder}
Here we summarize the formula for the largest amplitude component of transit timing variations of a pair of planets near a first order mean motion resonance, as derived by \citet{Lithwick2012}, and remind the reader of the origin of the mass-eccentricity degeneracy.

\citet{Lithwick2012} considered a system of two planets near the $j_R$:$j_R-1$ first order resonance, in which case the TTVs take the following approximate form:
\begin{eqnarray}\label{eqn:Lithwick2012A}
\delta t_1 & =&  \Re{[-iV_1 \exp{i \lambda^{j_R}}]} \cr
\delta t_2 & =&  \Re{[-iV_2 \exp{i \lambda^{j_R}}]};
\end{eqnarray}
here $i$ denotes $\sqrt{-1}$ and
\begin{eqnarray}\label{eqn:Lithwick2012B}
V_1 &=& \frac{P_1}{\pi}\frac{1}{j_R^{2/3}(j_R-1)^{1/3} \Delta}\frac{m_2}{M_\star}\bigg[- f -\frac{3Z_{free}^*}{2 \Delta}\bigg] \cr
V_2 &=& \frac{P_2}{\pi}\frac{1}{j_R \Delta}\frac{m_1}{M_\star}\bigg[- g +\frac{3Z_{free}^*}{2 \Delta}\bigg] \cr
\lambda^{j_R} &=& j_R\lambda_2-(j_R-1)\lambda_1,
\end{eqnarray}
with
\begin{eqnarray}\label{eqn:Lithwick2012C}
Z_{free} & = & f e_1^{i \varpi_1}+g e_2^{i \varpi_2}\cr 
\Delta &=& \frac{P_2}{P_1}\frac{j_R-1}{j_R}-1
\end{eqnarray}
and $f$ and $g$ are combinations of Laplace coefficients (note that $f<0$ and $g>0$), $e_{1,2}$ the free eccentricities, and $\varpi_{1,2}$ the longitudes of pericenter of each of the planets.  The reference direction from which longitudes are measured is the observer's line of sight, so that $\lambda_i=0$ at transit. Note that $V_1$, $V_2$, and $Z_{free}$ are complex quantities;  however, all observables are found by taking the real components as in Equation \ref{eqn:Lithwick2012A}.  ($Z^*_{free}$ is the complex conjugate of $Z_{free}$, and $\Re(x)$ and $\Im(x)$ denote the real and imaginary components of a complex number $x$). In the derivation of these formulae, only the two resonant terms associated with the $j_R$:$j_R-1$ resonance at first order in eccentricity are considered. All other terms in the gravitational potential between the planets, even those of independent of eccentricity, are neglected since near resonance the TTVs they produce are very small compared to those caused by the resonant terms (due to the small denominator $\Delta$ appearing in the amplitudes given in Equations \ref{eqn:Lithwick2012B}).

The total eccentricity of a planet near a first-order mean motion resonance is made up of ``free" and ``forced" components; the forced eccentricity is driven by resonant interactions between the planets while the free eccentricity is determined by matching the initial eccentricity vector (the initial conditions). The free eccentricity vector can be approximated as constant on observational timescales (it varies on the long secular timescale) while the forced eccentricity vector precesses on the resonant timescale (the super-period). Both components affect the TTVs: the forced eccentricity is responsible for the first term in the brackets in Equations \ref{eqn:Lithwick2012B} (depending on the coefficients $f$ and $g$), while the free eccentricities are responsible for the $Z_{free}^*$ term. The amplitude of the free eccentricity is not necessarily smaller or larger than that of the forced eccentricity. 

In the case of two transiting planets, the unknown quantities are $\Re[Z_{free}]$, $\Im[Z_{free}]$, $m_1/M_\star$, and $m_2/M_\star$, and, in principle, there are four measurable quantities: both amplitudes and both phases. However, as pointed out in \citet{Lithwick2012}, in both the case of $\vert Z_{free} \vert << \Delta$ and  $\vert Z_{free} \vert >> \Delta$ the TTVs are approximately anti-correlated, and so in either of these regimes the relative phase is no longer a constraining quantity (if the signal-to-noise is insufficient to measure a phase offset). This leads to the degeneracy between mass and free eccentricities, and hence from this first-order TTV alone one in practice cannot usually determine the masses or the combination of free eccentricities $Z_{free}$.

Finally, since the TTVs are sampled for the inner [outer] planet at $\lambda_1(t=t_{1,k}) = \lambda_1(t=t_{1,0})+2\pi k$ [$\lambda_2(t=t_{2,k}) = \lambda_{2}(t=t_{2,0})+2\pi k$], these expressions become: 
\begin{eqnarray}\label{eqn:Lithwick2012D}
\delta t_1 & =&  \Re{[-iV_1 \exp{(i j_R \lambda_2})]} \cr
\delta t_2 & =&  \Re{[-iV_2 \exp{(i (j_R-1)\lambda_1)}]},
\end{eqnarray}
Actually, Lithwick et al. have taken advantage of the aliasing effect to create these expressions. The component of the TTVs due to the forced eccentricity has a $j_R(\lambda_2-\lambda_1)$ dependence for the inner planet and a $(j_R-1)(\lambda_2-\lambda_1)$ dependence of the outer planet (see their Equations (A15) and (A24)). Because these appear at the same aliased frequency as the resonant $j_R\lambda_2-(j_R-1)\lambda_1$ term, they are included with the resonant TTV. Since the reference direction was chosen such that $\lambda_i=0$ at transit, these two terms have the same phase and can be grouped together in this way.

\section{The Synodic Chopping Signal}\label{sec:chopping}
A pair of planets interacts most strongly when the distance between them is smallest. For low eccentricity and nearly coplanar orbits, this occurs at conjunction, when the synodic angle $\psi = \lambda_1-\lambda_2=0$ ($\lambda_1=\lambda_2$).  Conjunctions occur periodically, with a timescale of $P_{syn} = 2\pi/n_{syn} = 1/(n_1-n_2) = 1/(1/P_1-1/P_2)$, and it is intuitive to expect that this timescale would appear in the TTVs. In fact, at zeroth order in the eccentricities, the transit timing variations only depend on the synodic angle and its harmonics,
\begin{equation}
\psi_j=j(\lambda_1-\lambda_2);
\end{equation}
see Appendix for more detail.

\subsection{Synodic chopping signal formulae to zeroth order in the free eccentricities} \label{sec:chopping_formula}

The synodic chopping signal is included in the computations in \citet{Agol2005} and
\citet{Nesvorny2014}. We also give a distinct derivation in the Appendix here which casts the expressions in a form useful for the harmonic analysis described in Section \ref{sec:harmonic}. All three formulae agree in the limit that the reflex motion of the star can be ignored (after correcting a typo in \citet{Nesvorny2014}; see Appendix for more detail).

For an inner transiting planet, the synodic component of the transit timing variations takes the form:
\begin{equation}\label{eqn:formulaInner}
\delta t_1 = \sum_{j=1}^\infty {P_1 \over 2\pi} {m_2 \over M_*} f_1^{(j)}(\alpha) \sin{ \psi_j},
\end{equation}
where
\begin{equation}
f_1^{(j)}(\alpha) = -\alpha \frac{j(\beta^2+3) b_{1/2}^{j}(\alpha)+2 \beta D_\alpha b_{1/2}^{j}(\alpha)-\alpha \delta_{j,1} (\beta^2+2\beta+3) }{\beta^2 (\beta^2-1)},
\end{equation}
where $\beta = j(n_1-n_2)/n_1$, $b_{1/2}^{j}(\alpha)$ is the Laplace coefficient
\begin{equation}
b_{1/2}^j(\alpha) = \frac{1}{\pi} \int_0^{2\pi} \frac{\cos{(j \theta)}}{\sqrt{1-2\alpha \cos{\theta}+\alpha^2}}d \theta.
\end{equation}
and $D_\alpha$ is the derivative operator $\alpha \frac{\partial}{\partial \alpha}$.

The Laplace coefficients can be evaluated in terms of complete elliptic integrals; for example:
\begin{eqnarray}
b_{1/2}^{(1)}(\alpha) &=& 4(K(\alpha)-E(\alpha))/(\alpha \pi), \cr
b_{1/2}^{(1)}(\alpha) + \alpha {\partial b_{1/2}^{(1)}(\alpha)\over \partial \alpha} &=& {4\alpha E(\alpha)\over \pi(1-\alpha^2)},
\end{eqnarray}
where $K(k)$ and $E(k)$ are the complete elliptic integrals of the first and second kinds, respectively.

For an outer transiting planet, the synodic component of the transit timing variations takes the form:
\begin{equation}\label{eqn:formulaOuter}
\delta t_2 = \sum_{j=1}^\infty {P_2 \over 2\pi} {m_1 \over M_*} f_2^{(j)}(\alpha) \sin{\psi_j},
\end{equation}
where
\begin{equation}
f_2^{(j)}(\alpha) =  \frac{j(\kappa^2+3) b_{1/2}^{j}(\alpha)+2 \kappa (D_\alpha b_{1/2}^{j}(\alpha)+b_{1/2}^{j}(\alpha))-\alpha^{-2} \delta_{j,1} (\kappa^2-2\kappa+3) }{\kappa^2 (\kappa^2-1)},
\end{equation}
where $\kappa = j(n_1-n_2)/n_2$.

In the limit of large $j$, the function $f_1^{(j)}(\alpha)$ has a leading coefficient which scales like $\alpha^{j+1}/j$ and the function $f_2^{(j)}(\alpha)$ has a leading coefficient which scales like $\alpha^{j+3}/j$  (see Appendix), so that the largest contributions to the sum in Equation \ref{eqn:formulaInner} and in Equation \ref{eqn:formulaOuter} in general come from smaller values of $j$. Because of the behavior of the leading coefficients, more terms are necessary to faithfully approximate the TTV signal as $\alpha \rightarrow 1$ because the convergence of the sums is slow.

Closely spaced planets will, at a given $j$, have a larger synodic TTV than more widely spaced planets  because the parameters $\beta$ and $\kappa$ appearing in the denominators approach zero as $\alpha \rightarrow 1$. Additionally, if the pair is near the $j_R:j_R-1$ mean motion resonance, the denominators $\beta^2-1$ and $\kappa^2-1$ will be close to zero for the term with  $j=j_R$ (for the inner planet) and $j=j_R-1$ (for the outer planet). Near the $j_R:j_R-1$ resonant configuration, then, these terms dominate the synodic TTV because of these small denominators. This reflects the fact that near the $j_R:j_R-1$ mean motion resonance, the time between conjunctions approximately corresponds to $j_R-1$ orbits of the outer planet and $j_R$ orbits of the inner one, such that most of the TTV amplitude is incorporated in these harmonics of the synodic angle. Additionally, these particular harmonics of the synodic angle contribute to the TTV as a long period effect. Due to aliasing, they have a timescale given by the super-period of the first order resonance.

\begin{figure}
\center
\includegraphics[scale=0.65]{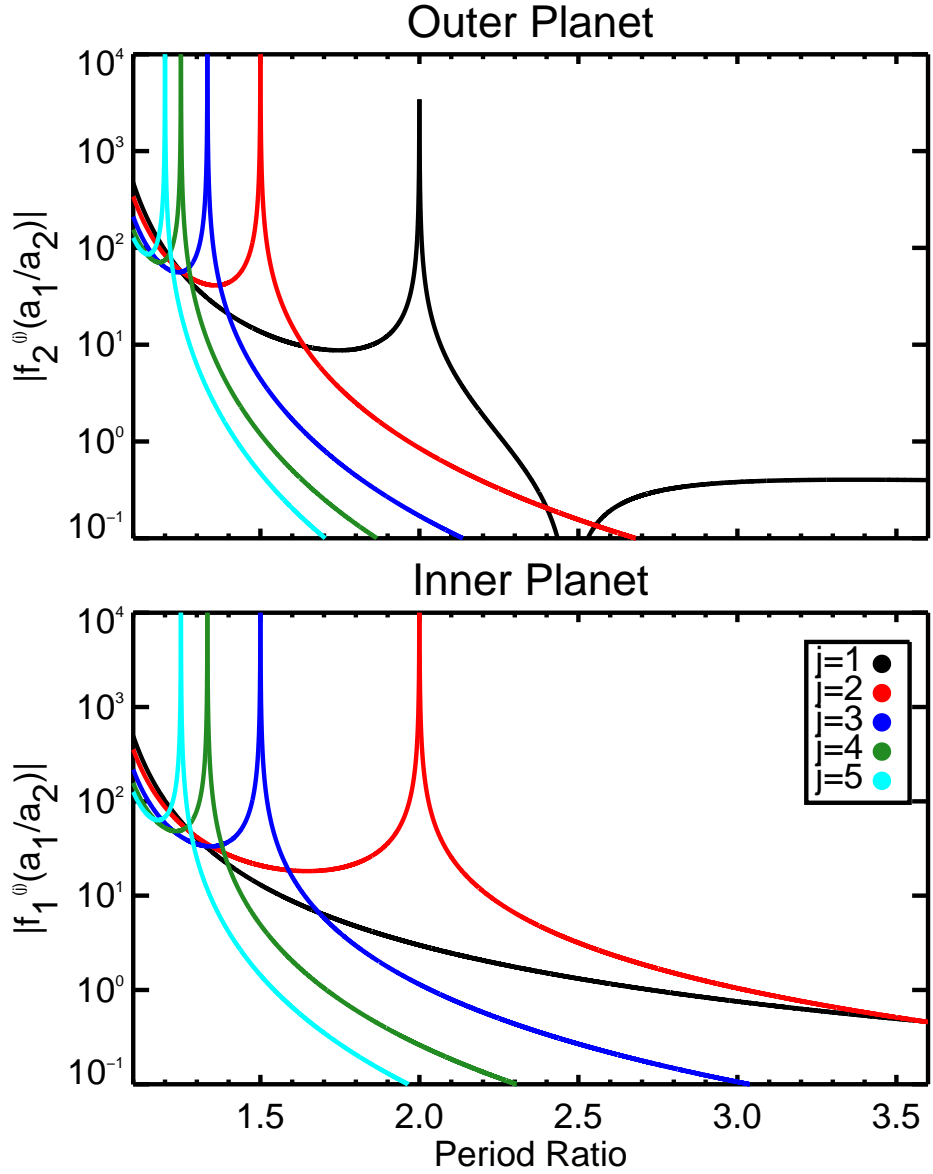}
\caption{Magnitude of the chopping coefficient functions, $f_1^{(j)}(\alpha)$ (bottom) and
$f_2^{(j)}(\alpha)$ (top), as a function of period ratio, for several values of $j$. Near the $j_R$:$j_R-1$ resonance, the inner planet chopping function peaks for $j=j_R$, while the outer planet chopping function peaks for $j=j_R-1$ (with $j_R\geq2$).}
\label{fig:chopping_j}
\end{figure}

The magnitude of the functions $f_1^{(j)}(\alpha)$ and $f_2^{(j)}(\alpha)$ are plotted in Figure \ref{fig:chopping_j} for $j={1,2,3,4,5}$. It is clear that the synodic amplitude $f_{1,2}^{(j)}(\alpha)$ generally grows as $\alpha \rightarrow 1$, for all $j$, and therefore more terms must be included in the sum for close pairs of planets. Furthermore, when the period ratio is close to the $j_R:j_R-1$ mean motion resonance, the functions with $j=j_R$ (inner planet) and $j=j_R-1$ (outer planet) peak for the reasons discussed above. This happens for all $j$ at some period ratio {\it except} for the $j=1$ synodic TTV of the inner planet since $j_R \geq 2$.  As noted in \citet{Agol2005}, the dip near a period ratio of 2.5 in the $j=1$ synodic signal of the outer planet is due to the fact that the TTV caused by  the motion of the star about the barycenter of the inner planet-star binary subsystem is opposite in sign and comparable to that caused by direct interaction with the inner planet at this period ratio. For larger period ratios, the TTV of the outer planet is dominated by the component due to the motion of the star about the barycenter of the inner planet-star binary subsystem.

Note that these relations have no eccentricity dependence; they only
depend on the semimajor axis ratio of the two planets, $\alpha$, and the planet-star mass ratios. 
The phase and period of the chopping signal are straightforward
to measure if the perturbing planet also transits the star; thus,
if this synodic chopping signal can be measured, the mass
ratio of the perturbing planet can be immediately inferred if the
parameters of the system satisfy the major assumptions made in this paper (see Section \ref{sec:EccEff}).

\subsection{Range of validity of the synodic formula}\label{sec:EccEff}
The intrinsic assumptions made in deriving these TTV expressions were 1) that the system has low eccentricities and nearly coplanar orbits; 2) that the system is not in or too near resonance; 3) that $\alpha$ is not too close to unity; 4) that the masses of the planets are small compared to that of the star; and 5) that the coefficients themselves can be treated as constant in time. Additionally, if two perturbing planets contribute to the TTVs of a third planet, the true TTV of this third planet cannot be written as only a sum of one of the perturbing planet's harmonics. In most configurations, we hypothesize that the TTV could be approximated as a simple sum of two ``single-perturber" contributions.

We will focus on testing the validity of the formula in light of the intrinsic sources of error in the formula. First, we consider the qualification of ``small" eccentricities and inclinations. A simple first guess of the order of magnitude error of these neglected terms would be that the coefficient would change from a value $C$ to $C[1+O(e)+O(i)]$, in which case if the eccentricities and inclinations are only a few percent than the error will also be a few percent. However, whether or not the eccentricities and inclinations can be considered ``small" depends on how close the system is to resonance. 

In general, if one derived a formula for the TTV good to first order in eccentricity, it would have terms at first order in eccentricity only depending on the angles of $j \lambda_2-(j-1) \lambda_1$ (not including the longitude of pericenter piece). In fact, the TTV formulae themselves should have the d'Alembert characteristics since they only depend on angles referenced to a fixed direction (longitudes) \citep{Hamilton}. Due to the transit aliasing effect discussed above, these first-order frequencies will be indistinguishable from variations of frequency $j(n_1-n_2)$ for the inner planet since these appear as the same harmonic of the outer planet, $jn_2$ (see Equation \ref{eqn:Lithwick2012D}). For the outer planet, these first order (in eccentricity) terms will indistinguishable from those at the harmonic of $(j-1)(n_1-n_2)$.

These first-order terms will have amplitude proportional to $e/(j n_2-(j-1)n_1)$, which, away from resonance, should be negligible for all $j$ if the eccentricity is low. But near the $j_R$:$j_R-1$ resonance, the $j_R n_2$ (for the inner planet) and the $(j_R-1)n_1$ term (for the outer planet) will have amplitudes proportional to $e/(j_R n_2-(j_R-1)n_1)$ (appearing as $e/\Delta$ in the formulae of Lithwick et al, Equation \ref{eqn:Lithwick2012B}). The resonant combination of frequencies in the denominator mitigates the effects of the eccentricity coefficient, so that these terms are large corrections to the TTV formulae at these frequencies, and hence they make the synodic formula for these values of $j$ less accurate. Note that since $j_R \geq 2$, the synodic frequency, $n_1-n_2$, has no resonant aliases for the inner planet. Thus the harmonic component which depends on $n_2$ can be uniquely identified with the synodic frequency variation.

If the system is {\it in} a mean motion resonance, the dominant, resonant TTV period will be related to the libration time. In this case, the resonant contribution to the TTVs will not appear at the frequency $j_R n_2$ for the inner planet and $(j_R-1) n_2$ for the outer planet, since the libration time cannot be written simply as an integer times the orbital frequency. In the resonant case, then, the harmonic analysis approach cannot be used. 

It is important to point out that in each of these cases - both near the $j_R$:$j_R-1$ resonance, where eccentricity errors are larger, and in the mean motion resonance, where the derivation in the Appendix must be modified, the synodic TTV formula for values of $j \neq j_R$ (inner planet) and $j\neq j_R-1$ (outer planet) will still apply.

\begin{figure}
\center
\includegraphics[scale=0.65]{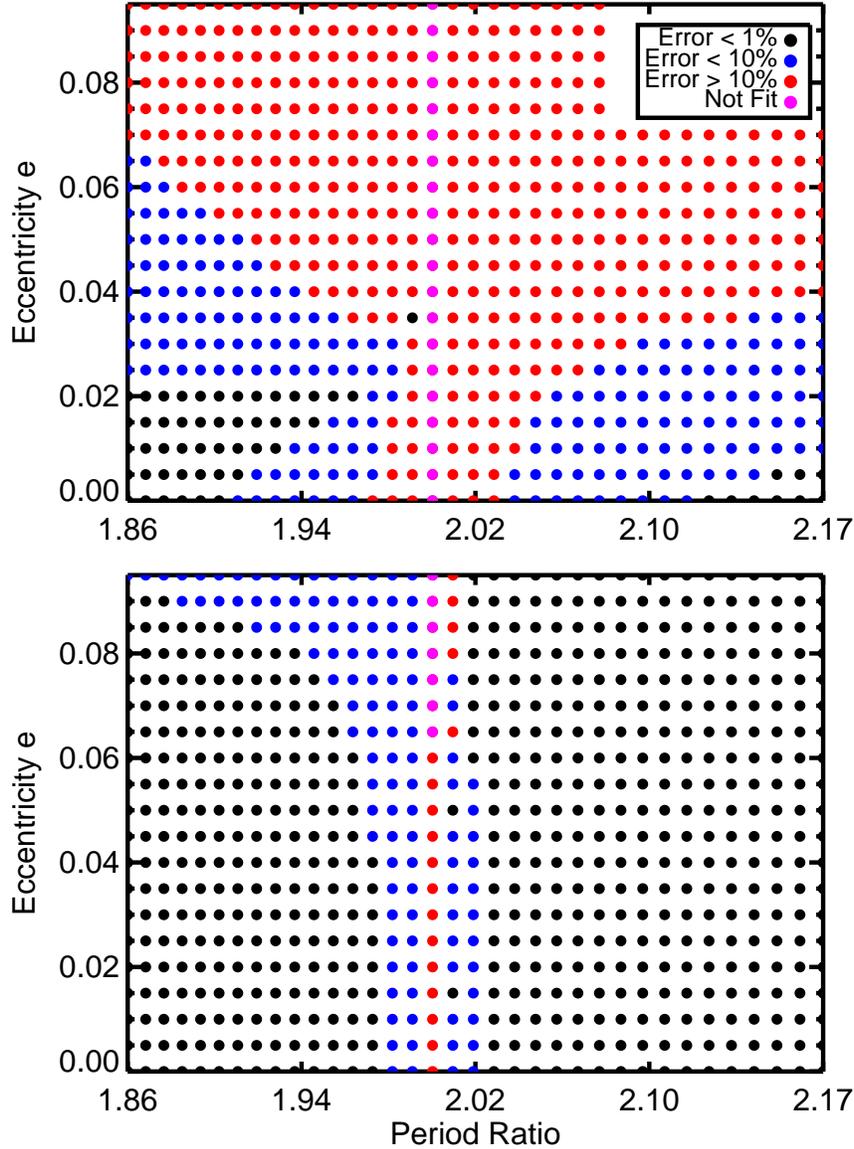}
\caption{Comparison of the synodic formula with $j=1$ with the measured harmonic with frequency $n_2$ of simulated TTVs for the inner planet (bottom), and with frequency $n_1$ of simulated TTVs for the outer planet (top). The color scale denotes the fractional error in the chopping formula as a function of the period ratio of the pair and the eccentricities of the orbits, where the vertical scale is $e =(e_1^2+e_2^2)^{1/2}$. Black points denote errors $E <1\%$, blue those with errors $1\%<E<10 \%$, and red those with errors $E>10 \%$. We did not fit the TTVs of the systems in magenta, because the fitting method failed (because the TTV period was significantly unresolved by the simulation time). For the outer planet we purposely picked the harmonic $j=1$ which is aliased to the resonant frequency in order to show how higher order eccentricity effects are important for understanding the resonant term.}
\label{fig:ComparisontoNumericsA}
\end{figure}

In Figure \ref{fig:ComparisontoNumericsA}, we show how the predicted amplitude of the $j=1$ synodic chopping term compares with that calculated numerically, for both the inner and outer planet, assuming the pair is near the 2:1 resonance. We varied only the period of the outer planet and the eccentricity of both planets (assumed to be equal; the vertical scale is the square root of the sum of the squares of the eccentricities, $e = (e_1^2+e_2^2)^{1/2}$). The mass of each planet was set to be $10^{-5}M_\star$. For each planet, we first determined the TTVs by fitting a line to the transit times, and we then used the computed average orbital periods to determine the value of $\alpha$ used in the synodic formulae with $j=1$ for each planet.  We then determined the numerical amplitude of the chopping signal by fitting 10 harmonics of the perturber's period along with a linear term to 1,000 simulated TTVs of the inner planet and between 460 and 540 simulated TTVs of the outer planet and selecting the $j=1$ harmonic. This experiment therefore represents an ideal case where the main source of error comes from the assumptions made in deriving the formulae, and not from issues with not having enough data and/or precision to resolve the amplitudes of the various harmonics.

The error in the formula for the inner planet is below $1\%$ across the entire range studied, except very near the 2:1 resonance itself. This is as expected -  near the 2:1 resonance, only the $j=2$ contribution of the TTV of the inner planet is expected to have large corrections due to eccentricity effects.  The configurations with errors of $>10\%$ are likely those in the mean motion resonance, where the harmonic approach does not apply (the libration frequency is not simply aliased with the frequency $2 n_2$), or those where the super period is significantly longer than the simulation time (to be discussed at the end of this section). For the outer planet, the $j=1$ synodic term is aliased with the $j_R=2$ resonant term, and hence we expect large errors as a function of eccentricity.   Indeed, even relatively far from resonance the synodic formula with $j=1$ fails (errors larger than 10\%) for eccentricities larger than 0.04.

We performed the same numerical experiment varying the number of harmonics fit to the TTVs, and we also decreased the simulation time. We found the same results even with fewer harmonics (in all cases, the number of data points was much larger than the number of free parameters). With shorter simulation times, we also recovered the same results except when the simulation time became significantly shorter than the TTV period (in which case the error is not due to the formula, but due to insufficient coverage).

Although we have not calculated the effects of inclination on the TTVs, if the TTVs in fact follow the d'Alembert characteristics then the correction due to inclination will only appear at second order. Therefore, one would expect that the inclination terms neglected will be a smaller source of error than the eccentricity terms neglected. Indeed, \citet{Nesvorny2014} did not find errors larger than 20\% in the synodic chopping formula (away from resonance) until the mutual inclination was larger than $50^\circ$.

In deriving the TTV formula, we began from the disturbing function, which assumes that the interaction between the two planets can be written as a converging series in $\alpha$. This means that the formulae will not work for co-orbital planets (see \citet{Nesvorny2014b} for an analysis of that case). Additionally, as $\alpha \rightarrow 1$, the Laplace coefficents converge less quickly, and so higher order eccentricity and inclination terms, ignored in the derivation, are potentially more important. A different issue is that as $\alpha \rightarrow 1$, mean motion resonances are densely spaced (for arbitrary $\alpha$, the system is more likely to be close to a resonance if $\alpha$ is closer to unity). In principle then the (neglected) effects of eccentricity could be more important for closer pairs of planets due to the effect of the small denominators discussed above. Note, however, again the special case of the $j=1$ synodic TTV of the inner planet. At $j=1$ there are no possible small denominators in the TTV of the inner planet, since all resonances (except the 1:1) require $j\geq 2$. 

\begin{figure}
\center
\includegraphics[scale=0.65]{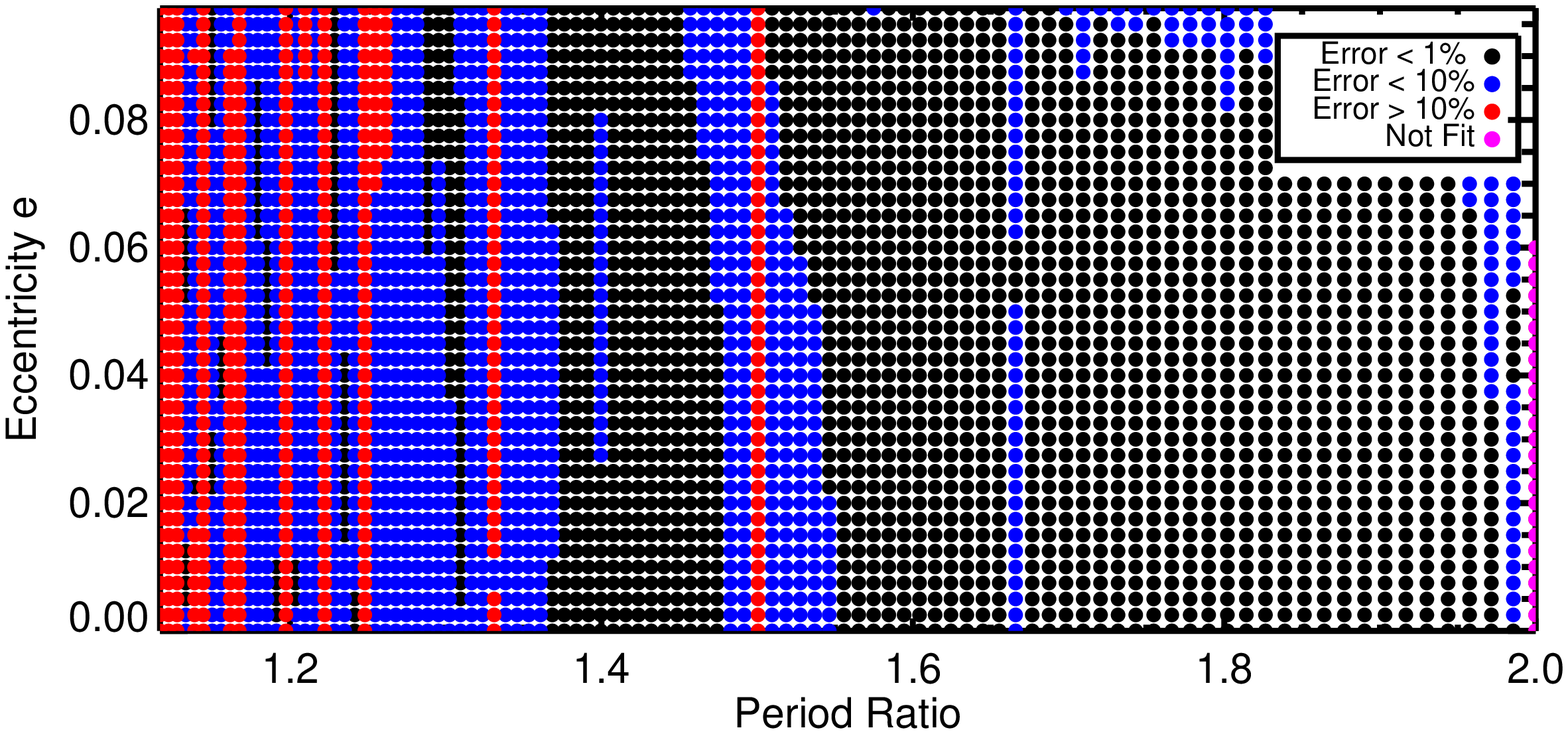}
\caption{Comparison of the synodic formula with $j=1$ with the measured harmonic with frequency $n_2$ of simulated TTVs for the inner planet. The color scale denotes the fractional error in the chopping formula as a function of the period ratio of the pair and the eccentricities of the orbits, where the vertical scale is $e =(e_1^2+e_2^2)^{1/2}$. Black points denote errors $0\%<\vert E \vert <1\%$,  blue those with errors $1\%< \lvert E\lvert <10 \%$, and red those with errors $\lvert E \rvert>10 \%$. Again we did not fit the TTVs of the systems in magenta.}
\label{fig:ComparisontoNumericsB}
\end{figure}

In Figure \ref{fig:ComparisontoNumericsB}, we show the error in the formula for the $j=1$ synodic TTV, for the inner planet, across a wide range of orbital separation and eccentricities.  In making this plot, 500 transits of the inner planet were simulated and 10 harmonics were fit to the resulting TTVs. Except at resonance, the error is less than $10\%$, regardless of the eccentricity. This indicates that the neglected eccentricity terms are in fact small across a wide range of $\alpha$. Note that the error is unbiased, in that it is not typically positive or negative.

\begin{figure}
\center
\includegraphics[scale=0.55]{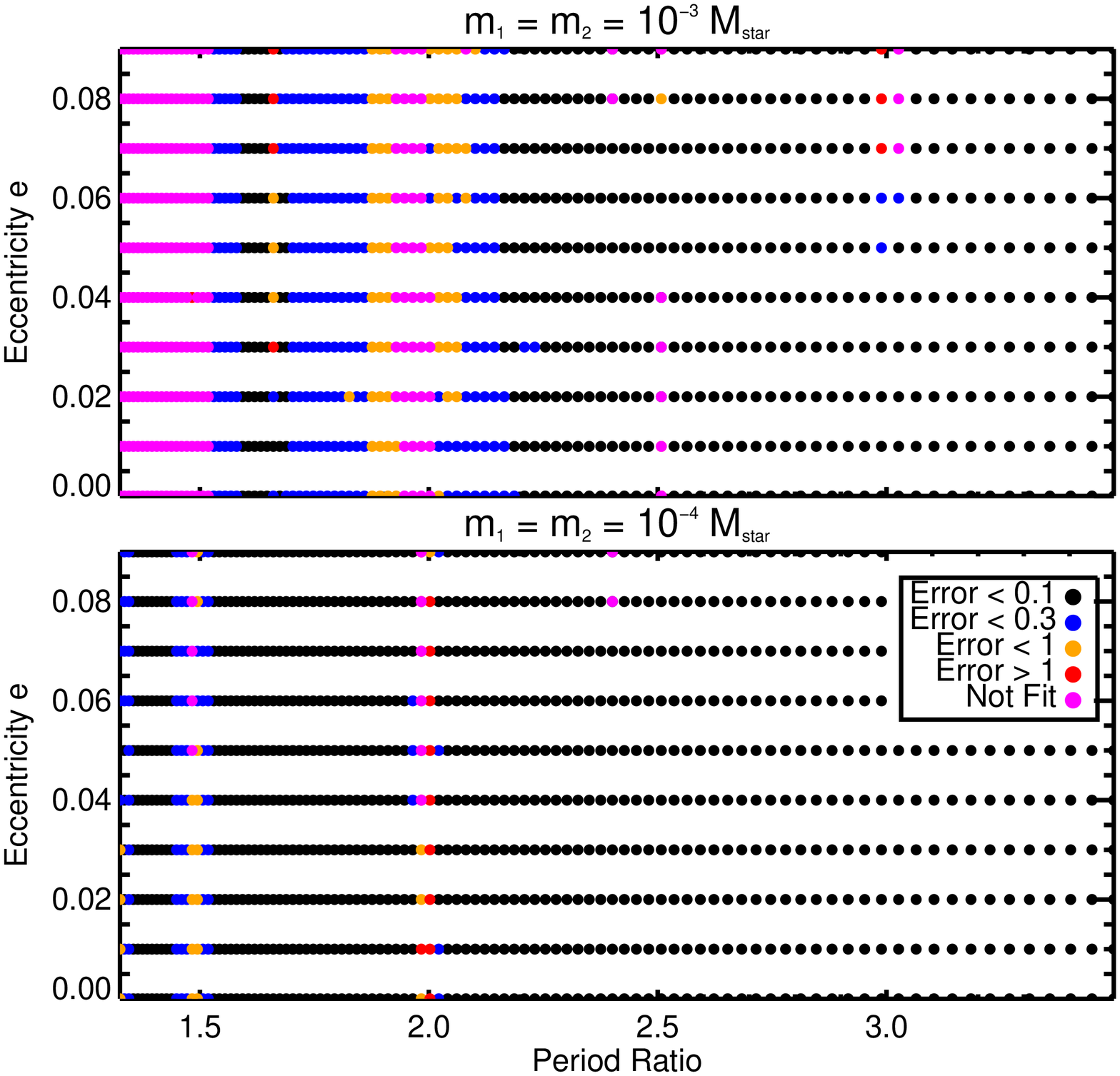}
\caption{Comparison of the synodic formula with $j=1$ with the measured harmonic with frequency $n_2$ of simulated TTVs for the inner planet for planets of mass $10^{-4}M_\star$ (bottom) or $10^{-3}M_\star$ (top). The color scale denotes the fractional error in the chopping formula as a function of the period ratio of the pair and the eccentricities of the orbits, where the vertical scale is $e =(e_1^2+e_2^2)^{1/2}$. Points in magenta had strongly perturbed orbits, with the number of transits varying by at least a few from the expected value, or else had TTV period significantly longer than the simulation time, and were not fit.}
\label{fig:masses}
\end{figure}

In deriving the synodic formulae, we assumed that the masses of the planets, compared to that of the host star, are small, so that neglected corrections of order $(m_{planet}/M_\star)^2$ are small. In Figure \ref{fig:masses}, we show how the fractional error in the synodic TTV formula with $j=1$ for the inner planet grows as we increase the masses of the planets to $10^{-4}M_\star$ and then to $10^{-3}M_\star$. Some of these configurations tested were unstable, and some were perturbed enough to change the number of transits by more than 1, and those were not fit (though shown in magenta). Note that the widths of resonances grow as the mass of the planets grow, and so more systems are in resonance, where the harmonic analysis will not work, than in the fiducial $m_{planet}/M_\star=10^{-5}$ case shown in Figure \ref{fig:ComparisontoNumericsB}. In the case of Jupiter mass planets, there are configurations where the chopping formula may be too approximate, as even outside of resonance, between the 3:2 and the 2:1, the error is on the order of $30\%$. As \citet{Nesvorny2014} point out, in these cases the chopping formula primarily provides motivation and understanding as to how mass measurements from TTVs arise.

Lastly, we assume that the coefficients of the synodic TTV, which depend on $\alpha$, can be treated as fixed in time at their average values.  It is possible, especially in the near resonant case where the TTV timescale is long, that the observations will cover only a small fraction of the TTV cycle. In this case, the observed semimajor axis ratio may be different than the average one. The fractional error resulting from using the ``incorrect'' value of $\alpha$ in the formula for $f^{(j)}_i(\alpha)$ will be of order $[1/f^{(j)}_i(\alpha)]  [\partial f^{(j)}_i(\alpha)/\partial \alpha] \delta\alpha$.  Although $\delta\alpha$ is small, of the same order of the TTV compared to the orbital period, the derivative of the coefficient $ f^{(j)}_i(\alpha)$ can be large near resonance. Therefore, if one observes only a small fraction of the super-period of a near resonant system, there will be errors relating to an incorrect estimate of the average value of $\alpha$. These errors are larger for the terms in the synodic sum aliased with the resonant frequency.

The long-period oscillations in the eccentricity and inclinations due to secular effects will not be resolved, since the secular timescale is in general very long compared to observational timescales (for example, on secular timescales $Z_{free}$ will change). The synodic chopping formula, which is independent of eccentricity and inclination, will therefore have slowly varying error terms, but as long as the eccentricities and inclination remain small over the secular timescale the error terms will remain small as well.

\subsection{Comparison to the Lithwick et al. formula with $Z_{free}=0$}\label{sec:compareLithwick}

How do the synodic chopping formulae compare with the Lithwick et al. formulae? First, the Lithwick et al. formulae applies for pairs of planets near a first order mean motion resonance, and it includes the contribution to the TTV at the frequencies $j_R n_2$ for the inner planet and $(j_R-1)n_1$ for the outer planet. However, since only the resonant terms were used in deriving the Lithwick et al. formulae, some synodic effects which get aliased to the resonant frequencies were neglected (these neglected terms are small amplitude near resonance since they do not have the small denominator of $j_R n_2-(j_R-1)n_1$).  On the other hand, the synodic formula applies for pairs both far and near mean motion resonances, and encompasses the effects of conjunctions at every harmonic $j n_2$ in the TTVs of the inner planet and $j n_1$ in the TTVs of the outer planet. However, the Lithwick et al. formulae includes the approximate first order eccentricity correction for pairs near a mean motion resonance, while the synodic formulae only hold at zeroth order in eccentricity. 

\begin{figure}
\center
\includegraphics[scale=0.55]{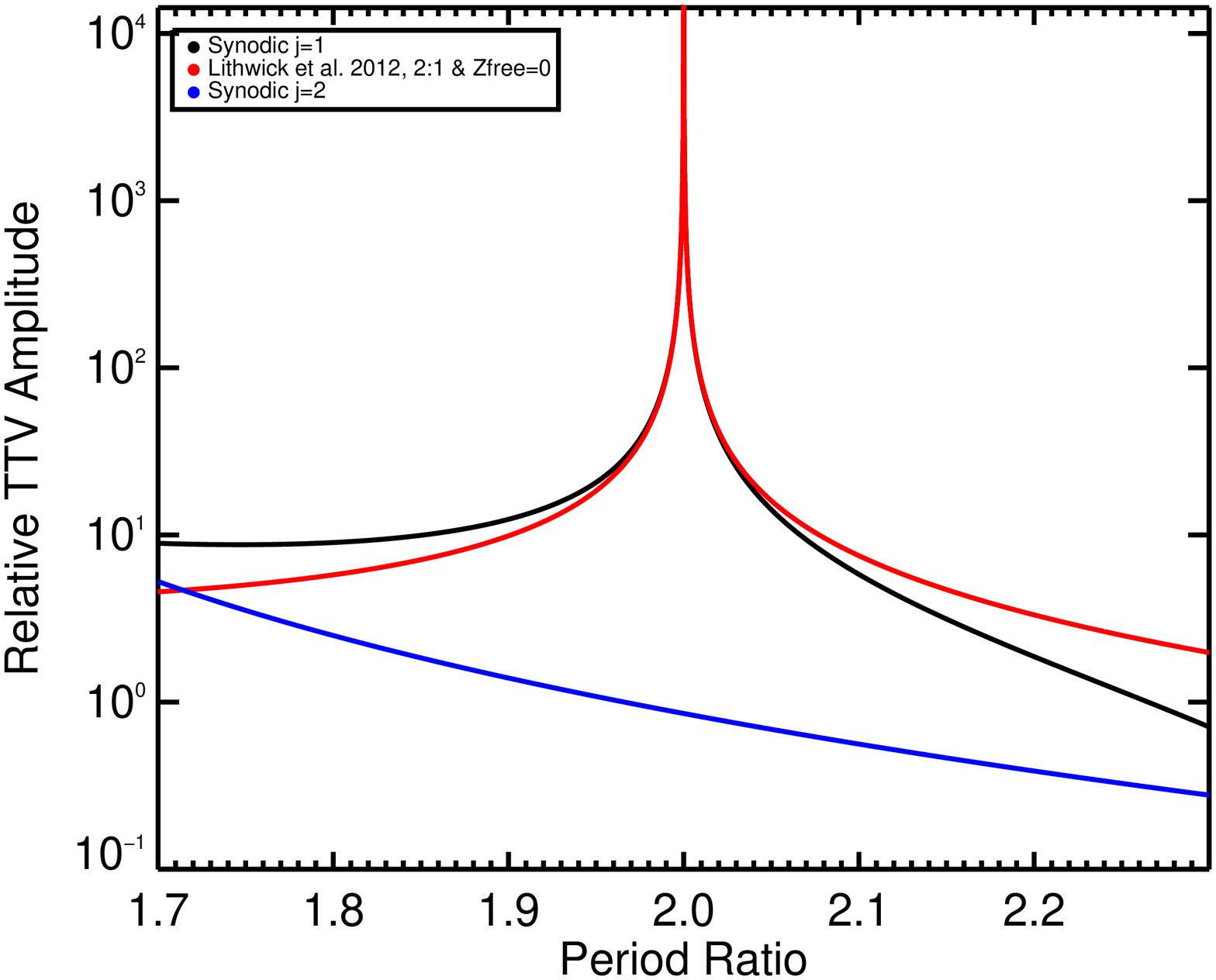}
\caption{Comparison of the amplitude of the synodic TTV of the outer planet, with $j=1$, and the amplitude from the TTV formula of Lithwick et al. 2012 with $Z_{free}=0$. Both the close agreement near the resonant period ratio of 2:1 and the disagreement further away is expected. For reference, the amplitude of the $j=2$ synodic component, which would be less affected by the eccentricity corrections important for near resonant systems, is also shown. }
\label{fig:comparison}
\end{figure}

We now compare the two sets of formulae in the regime where they should agree: near resonance (where the synodic chopping terms without the small denominator ignored by \citet{Lithwick2012} are negligible), with $Z_{free}=0$, and for the correct value of $j$ chosen in the synodic sum (that of $j=j_R$ for the inner planet and $j=j_R-1$ for the outer planet). Our expectation is borne out by a numerical comparison between the two, the results of which are shown in Figure \ref{fig:comparison}. In short, the Lithwick et al. expression is an excellent approximation to the TTV of systems near first order mean motion resonances, while further away from resonance it becomes a worse approximation to the chopping signal with $j=j_R$ for the inner planet and $j=j_R-1$ for the outer planet. Correspondingly, the synodic formulae will be a worse approximation at these specific values of $j$ for systems closer to a first order resonance because of eccentricity effects, as we neglect the $Z_{free} \neq 0$ correction - see Section \ref{sec:EccEff}.

\subsection{Measurement precision}\label{subsec:predict}
One can use the synodic TTV formulae to estimate the precision on the mass measurement of a perturbing planet. For example, the expected mass precision (of the outer planet) due to synodic chopping with $j=1$ in the inner planet is given by 
\begin{equation}\label{eqn:precision}
\sigma_{M_2} = \sigma_{t_1} M_* {2\pi \over \sqrt{(N_{trans}-N_{param})/2} P_1 f^1_1(\alpha)},
\end{equation}
where $\sigma_{t_1}$ is the timing precision of the inner planet,
$N_{param}$ is the number of model parameters, and $N_{trans}$ is the
total number of transits observed. This formula assumes that the
phase of the sine function is adequately sampled so that the RMS
of $1/\sqrt{2}$ can be assumed, that there is no uncertainty on the mass of the star, $P_1$, or $\alpha$, and that there is no covariance
with other harmonics being fit.  This formula also assumes that
transits are observed continuously over the full super-period.

\section{Applications}\label{sec:applications}
In this section we apply the chopping and resonant formulae to several planetary systems which have been analyzed using a full dynamical model involving numerical integration of the gravitational equations of motion. We instead analyze these systems using the synodic formulae for TTVs, in combination with the $Z_{free} \ne 0$ component of the Lithwick et al. formula for near resonant systems. As a result, we can measure planetary masses for these systems without full numerical analysis and without the complication of the mass-free eccentricity degeneracy inherent in the Lithwick et al. formula alone. This allows us to demonstrate empirically the validity of the synodic TTV signal and to strengthen the understanding of what information in a TTV signal leads to mass measurements.

For a system far from a first order resonance, we would recommend using the synodic formulae for the $j\psi$ terms (for which $Z_{free}=0$), and using the second component of the Lithwick et al. formula for the $Z_{free}\ne 0$ term for the resonant $j_R n_2$ for the inner planet and $(j_R-1)n_1$ for the outer planet. If the system is close to resonance, the synodic chopping formulae can be used for values of $j$ not aliased with the resonant frequencies. 
Note again that the resonant term and the synodic terms aliased with the resonant term will in general have different phases, unless the reference direction is chosen so that $\lambda=0$ at transit. Additionally, while only the sinusoidal component of the $j(\lambda_1-\lambda_2)$ harmonics should have nonzero amplitude, the resonant component $j_R \lambda_2 -(j_R-1)\lambda_1$ has in general both sine and cosine components due to the complex quantity $Z_{free}$.

\subsection{PH3/Kepler-289}

The Kepler-289 (PH3/KOI-1353/KIC 7303287) planetary system was identified by the Kepler pipeline \citep{Borucki2011,Batalha2013,Tenenbaum2013} and by the Planet Hunters crowd-sourced project \citep{Fischer2012}. This system consists of three planets with orbital periods near 35 days (Kepler-289b/PH3b), 66 days (PH3c), and 126 days (PH3d);  each adjacent pair of planets is close to a period ratio of 1:1.9 \citep{Schmitt2014}.  The outer two planets both display
large amplitude transit-timing variations with a timescale of the super-period of the nearby 2:1 resonance, and the middle planet shows
a strong chopping signal caused by the outermost planet.  The masses of the outer two planets were measured by the transit timing variations through a full numerical analysis of the (assumed coplanar) system, performed by EA, as part of \citet{Schmitt2014}. The inner planet does not have significantly detected TTVs, and does not significantly
affect the outer two planets' transit times, resulting in an upper
limit on its mass only.

\begin{figure}
\center
\includegraphics[scale=0.65]{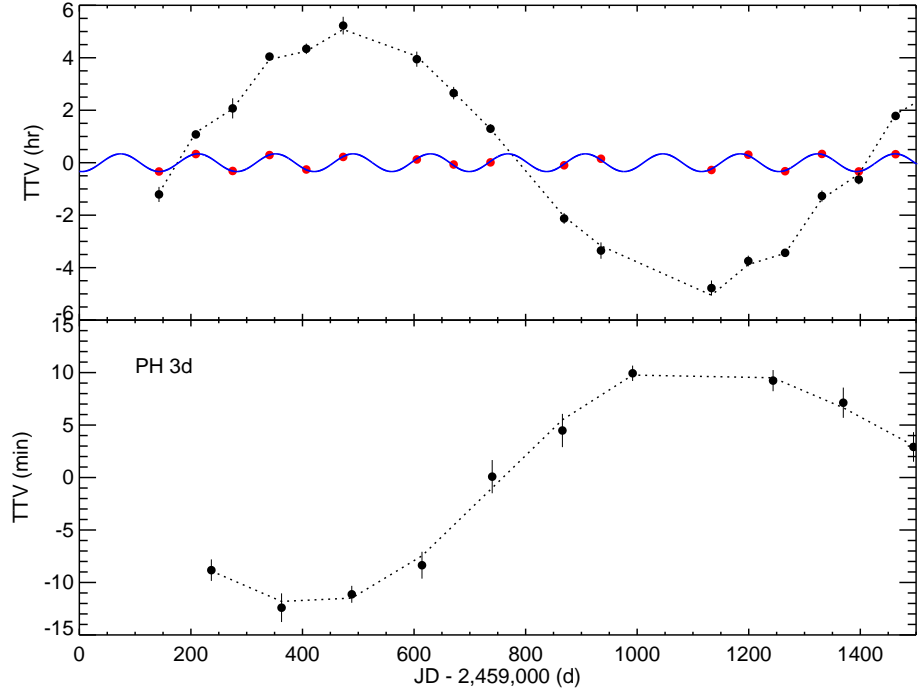}
\caption{Results of a harmonic analysis for PH 3c (top) and 3d (bottom). In black circles we show the full TTVs for each planet with the measurement uncertainties; note that the top panel is in units of hours, while the bottom is in units of minutes. The dotted curves are the best fit to the TTV using a harmonic analysis with two (one) harmonics of the perturbing planet's orbital frequency for PH 3c(d). In red is the measured chopping ($j=1$) component for PH 3c. The blue solid curve shows the predicted chopping signal based on the \citet{Schmitt2014} mass ratio for the outer planet.
}
\label{fig:koi1353_chopping}
\end{figure}

In this work, we returned to the published transit times and uncertainties of \citet{Schmitt2014} and analyzed them using the harmonic-fitting approach described in section \ref{sec:harmonic} in order to measure the masses of PH3c and PH3d.  Figure \ref{fig:koi1353_chopping} shows our initial harmonic fit to the data;  two harmonics are required for PH3c, while only one is required for 3d.  We ignore
the innermost planet (PH3b) in the analysis. We used the $Z_{free}\ne 0$ component of the Lithwick et al. formula with $j_R=2$ for modeling the $n_1$ component of the outer planet's TTVs and for the $2n_2$ component of the middle planet's TTVs.  We included the synodic chopping signal with $j=1,2$ for the middle planet PH3c,
and the $j=1$ component for the outer planet (which encompasses the $Z_{free}= 0$ contribution from the Lithwick et al. formula).

As these three terms together only constrain a linear combination of
the free eccentricities of the planets ($Z_{free}$), we also enforced as a prior the Hill stability criterion \citep{Gladman} to prevent the eccentricities from growing
too large.  We added a systematic error parameter, $\sigma_0$,
in quadrature to the measured timing errors, such that larger values of $\sigma_0$ are penalized in the likelihood function while smaller values of $\sigma_0$ require a closer fit to the transit times in order to have a high likelihood. We carried out an affine-invariant Markov chain analysis 
\citep{ForemanMackey2013} with eleven free parameters: the 
ephemerides ($t_0$, $P$), eccentricity vectors ($e \cos{\omega}$,
$e \sin{\omega}$), and masses of each planet, plus $\sigma_0$.

\begin{figure}
\center
\includegraphics[scale=0.85]{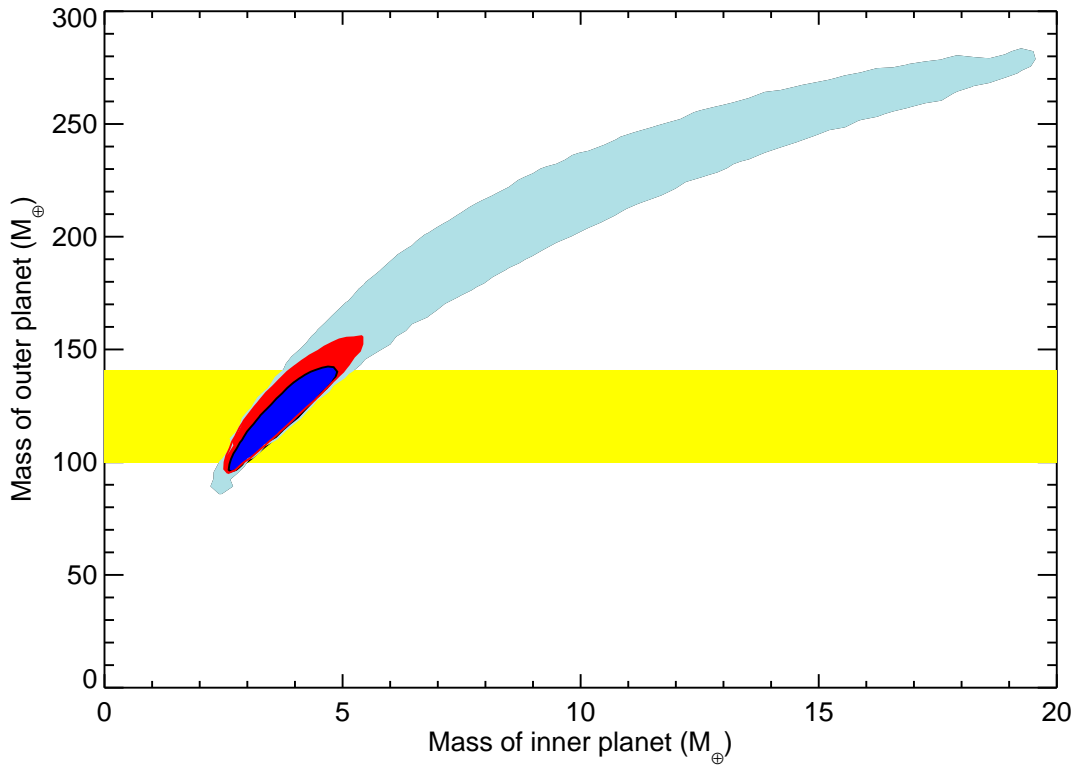}
\caption{Derived 1-$\sigma$ confidence limits for PH3c and d.
The light blue is derived from the harmonic analysis without fitting
the synodic chopping signal.  The red confidence limit includes
synodic chopping, while the blue shows the results from the
full dynamical analysis in \citep{Schmitt2014}. The constraint on the mass of the outer planet from the chopping signal alone is marked by the yellow region. } 
\label{fig04}
\end{figure}

Figure \ref{fig04} shows the
confidence limits in the planet masses from the harmonic analysis with and without the constraint from the synodic
chopping signal, as well as the confidence limits from the full dynamical analysis of \citet{Schmitt2014}.  As explained in \citet{Lithwick2012},
the 2:1 resonant signal constrains a combination of the mass ratios
of the planets and the free eccentricity, $Z_{free}$.  Without
the synodic chopping signal, our analysis shows a banana-like degeneracy
between the two planet masses which is due to the trade-off between their
masses and the free eccentricity (Figure \ref{fig04}, light blue), giving $M_c = 12 \pm 10 M_\oplus$ and $M_d=219 \pm 57 M_\oplus$. 
When the $j=1$ chopping signal in the TTVs of the middle planet is included,
the mass ratio of the outer planet becomes constrained;  this
then breaks the mass/free eccentricity degeneracy, and allows the
mass of the inner planet to be determined as well.  The
derived error ellipse is similar to that from the full dynamical analysis:
\citet{Schmitt2014} report masses of $M_c  = 4.0\pm 0.9 M_\oplus$
and $M_d = 132 \pm 17 M_\oplus$, while the harmonic analysis yields
$M_c = 4.3 \pm 1.1 M_\oplus$ and $M_d = 140 \pm 17 M_\oplus$.

This analysis demonstrates the power of the chopping signal
in constraining planetary masses near a first-order mean motion resonance. In principle, although the $j\geq2$ chopping components (for the outer planet) and $j\geq 3$ chopping components also provide independent constraints on the planetary masses, they are smaller in amplitude and not detected in this case. 

\subsection{Kepler 11d/e}

The Kepler-11 system \citep{Lissauer2011} is a system with six transiting planets.   A full dynamical
analysis has been carried out for this system, giving constraints
on the masses of all planets \citep{Migaszewski2013, Lissauer2013}.
Several of the planets, despite being only a few Earth masses, have
low densities which require H/He atmospheres;  this result is puzzling in light of core-accretion theory, which would not predict planets so low in mass to accumulate substantial gaseous envelopes. Here we validate the existing mass measurements of these two planets and we show that the mass constraints of Kepler-11d and Kepler 11-e largely
result from the chopping TTV signal.

Kepler 11d and 11 e are the two of the three most massive planets in
the Kepler-11 system, with periods near 23 and 32 days, respectively,
in close proximity to 3:2 commensurability. 
Each of these planets have transit timing variations that are dominated 
by the other;  thus they can be dynamically `decoupled' from the rest
of the planets, and treated as a two-planet system. Note, however, that the decoupling is ``one-way'': Kepler 11e affects the TTVs of planet Kepler 11-f, and hence there is more information with regards to the masses of d and e to be gained by fitting the entire system instead of treating the (d,e) pair in isolation.

\begin{figure}
\center
\includegraphics[scale=0.75]{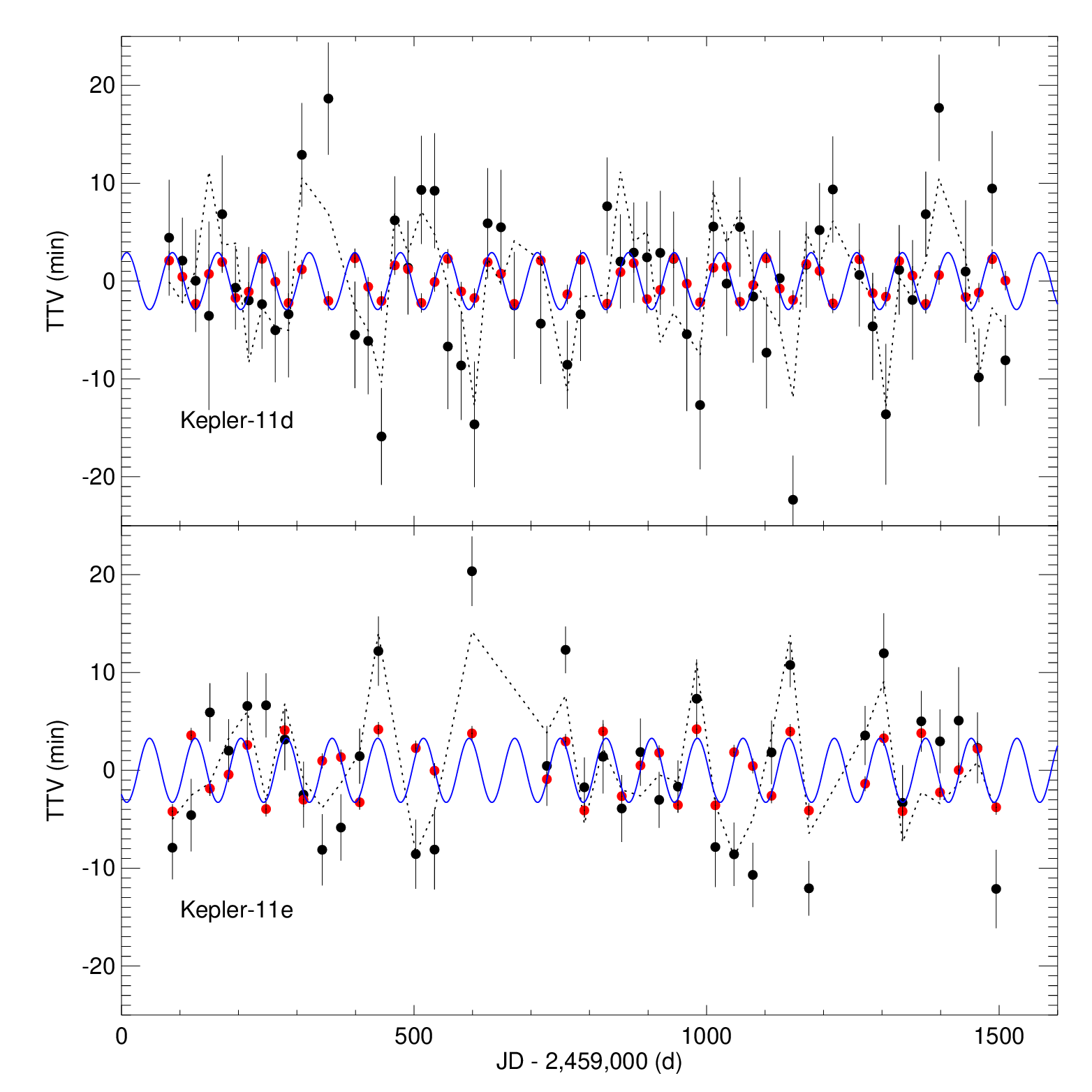}
\caption{Results of a harmonic analysis for Kepler 11d (top) and 11e (bottom). In black circles we show the full TTVs for each planet with the measurement uncertainties. The dotted curves are the best fit to the TTV using a harmonic analysis with three harmonics of the perturbing planet's orbital frequency. In red is the predicted chopping ($j=1$) signal for each planet. The blue solid curve shows the predicted chopping signal based on the \citet{Lissauer2013} results.
}
\label{fig:kepler11de_chopping}
\end{figure}

Figure \ref{fig:kepler11de_chopping} shows the harmonic fitting results
for Kepler 11d/e using the transit times due to Jason Rowe presented
in \citet{Lissauer2013}, to be compared to the dynamical constraints in
Table 7 of that paper. In black circles, we show the actual TTV measurements for 
each planet, with corresponding uncertainties. In this case we simply fit for the harmonics
of the TTV with the frequency of the companion planet up to $j=3$ (dotted lines), 
which resulted in excellent $\chi^2$ fits for both;  the $j=1$ 
synodic chopping signal from the harmonic fit is also plotted (in red). Note that near the 3:2 mean motion resonance, the $j=1$ synodic signal is not aliased with the resonant frequencies, and so we expect the chopping signal to be well approximated by our formula, as discussed in \ref{sec:EccEff}.  We over-plot
the predicted $j=1$ synodic chopping signal based on the the best-fit mass
ratios from Table 7 in \citet{Lissauer2013}, shown as the blue curves. This shows that the chopping signal is 
detected for both planets, and that it is consistent with the chopping signal predicted by the full dynamical 
analysis.

\begin{figure}
\center
\includegraphics[scale=0.65]{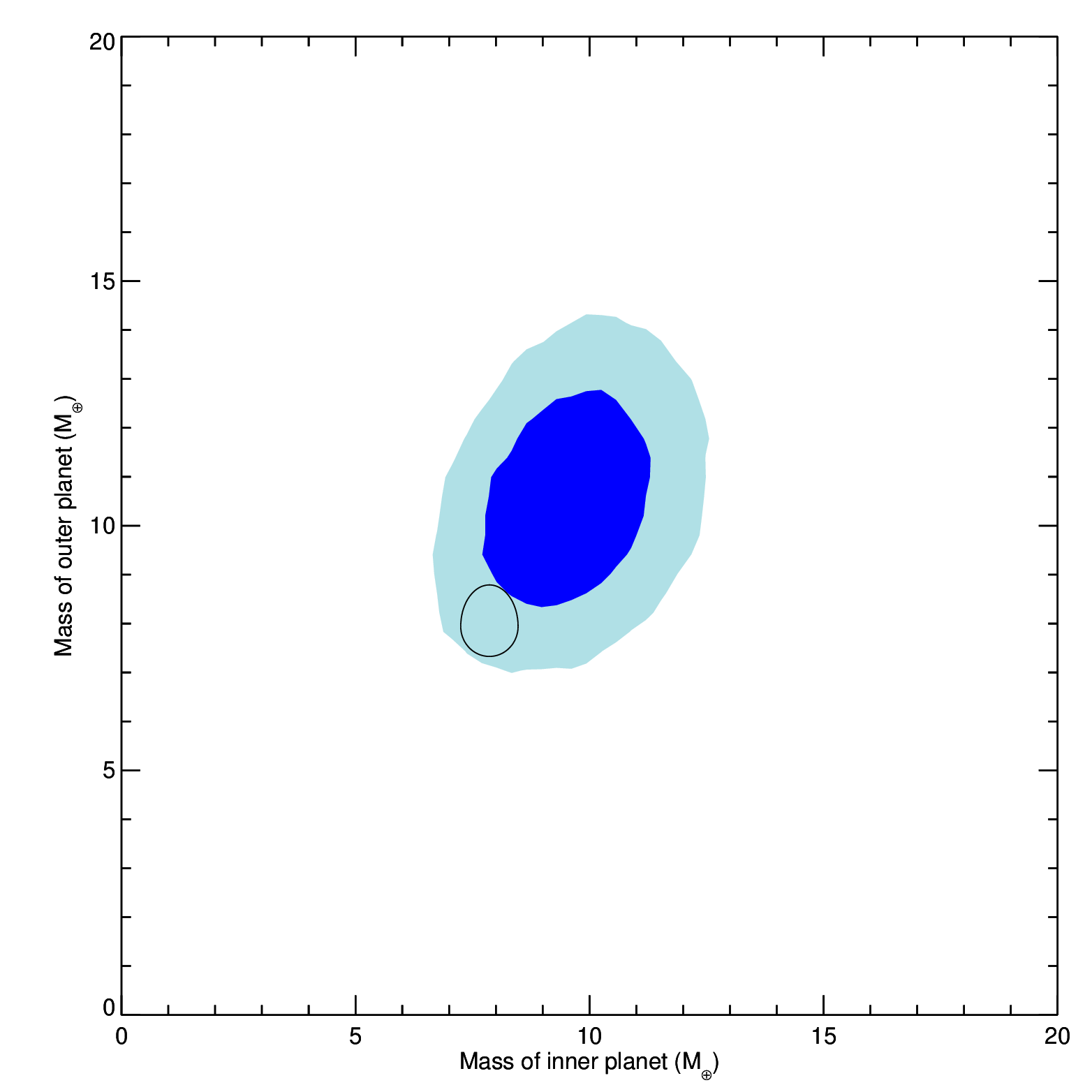}
\caption{Comparison of Kepler-11d/e resonant + chopping
analysis.  Black curve: 1$\sigma$ constraint from
Lissauer et al. (2014).  Dark[light] blue: 1[2]$\sigma$ constraint
from 3:2 resonant term and chopping terms for both planets.}
\label{fig:kepler11de_posterior}
\end{figure}

Next, we carried out a Markov chain analysis for these planets 
including the $Z_{free}\ne 0$ Lithwick et al. resonance
formulae with $j_R=3$, relevant to the 3:2 commensurability (Section 
\ref{sec:firstorder}), as well as both the inner and outer chopping 
formulae, summed to $j=4$.
Figure \ref{fig:kepler11de_posterior} shows the constraints on the masses
of the two planets. 
The black curve shows the 1$\sigma$ confidence limit from 
dynamical analysis in \citet{Lissauer2013}.
In dark[light] blue is the 1[2]$\sigma$ confidence limit for our analysis 
with the resonant and full chopping signals included for both
planets;  this is consistent with the dynamical analysis at
the $<2\sigma$ level, albeit with a larger uncertainty (recall again that the 
TTVs of planet Kepler 11-f are affected by Kepler 11-e, and so there is more information as to the (d,e) subsystem, available when fitting the whole system).

Another interesting byproduct of the chopping signal, when using in conjunction with the Lithwick et al. formula, is that it allows for a measurement of the quantity $Z_{free}$. In the case of Kepler 11 d/e, we find a value of $Z_{free} = (0.054\pm 0.017) +i(0.053 \pm 0.014)$. Since dissipation of eccentricities first damps the free eccentricities, it is interesting that the $Z_{free}$ here, though modest, is distinctly nonzero. 

In general, because the chopping amplitude function $f_i^j(\alpha)$ is smaller in magnitude for larger values of $j$, the components with larger $j$ may not be measurable. However, if they are, they can be used to provide additional constraints on the masses as consistency checks. For example, in the case of Kepler 11 d, the chopping signals with $j=1,2$ all independently constrain the mass of planet e, while the chopping signals of $j=1,3$ present in the TTVs of e all independently constrain the mass of planet d. For planet d, the inferred mass (assuming in this case a one solar mass star) is $10.4 \pm 1.8 M_\oplus (j=1)$ or $11.1 \pm 2.2 M_\oplus (j=3)$, compared to $7.86\pm0.61 M_\oplus$ in \citet{Lissauer2013}.  For planet e the inferred mass is $7.5 \pm 2.9 M_\oplus (j=1)$ or $10.5\pm 2.3 M_\oplus (j=2)$,
compared to $7.94\pm0.85 M_\oplus$ in \citet{Lissauer2013}.  Given that these estimates agree at the $(1-2)\sigma$ level, this indicates that the dynamical analysis yields masses that are consistent with the chopping amplitudes.

\subsection{Kepler 9}
In some cases, the Lithwick et al. formula alone can be used to determine the masses of the planets. In practice this requires that the amplitudes and phases of the TTVs must be measured with high accuracy.

\begin{figure}
\center
\includegraphics[scale=0.6]{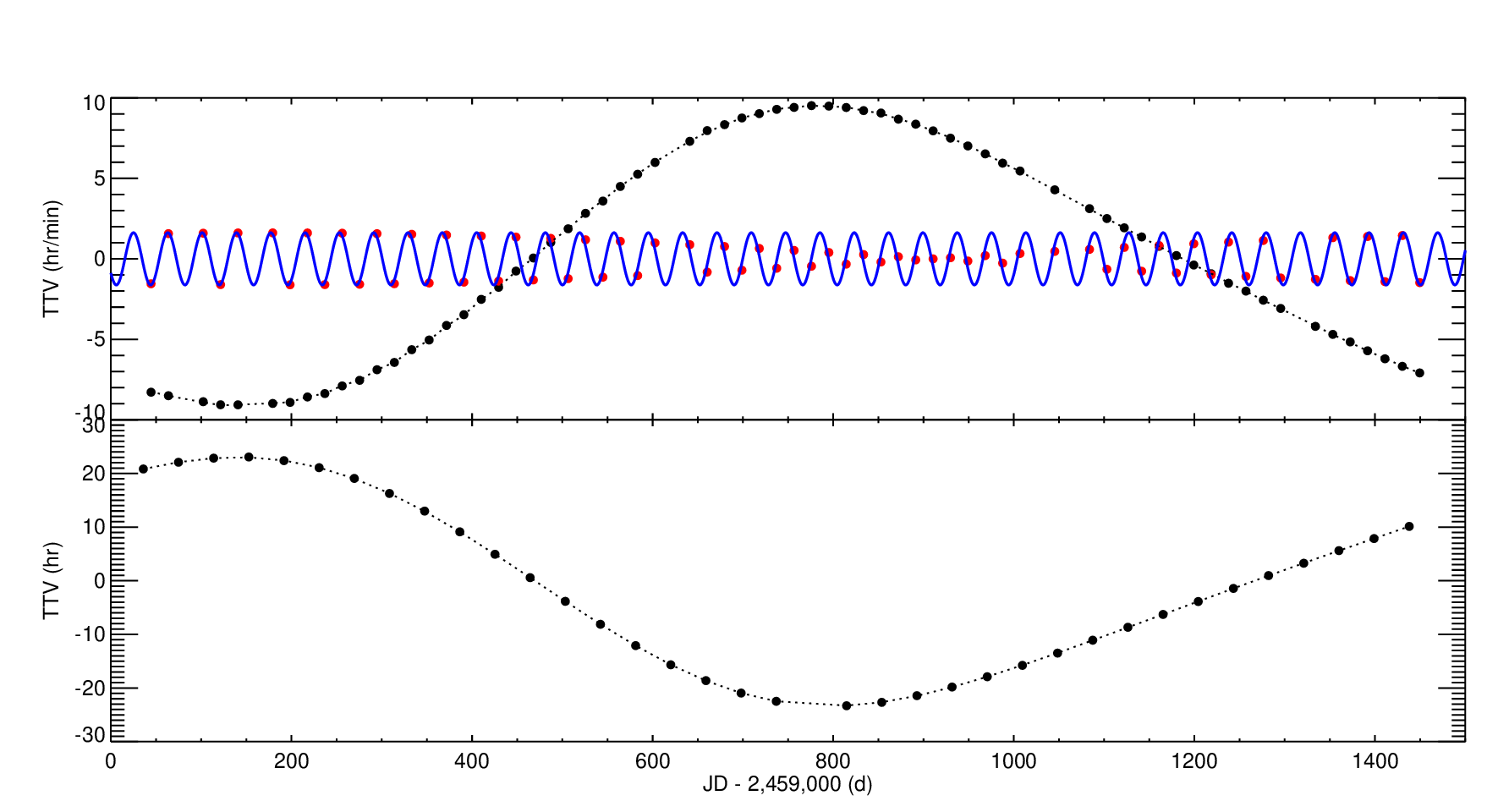}
\caption{Transit timing variations observed with Kepler for
Kepler-9 (black filled circles), in units of hours.  We have fit
six (four) harmonics to the transit times of Kepler-9c(d), 
obtaining an excellent fit (dotted line).  The synodic chopping 
component of our fit to the inner planet is
plotted with red dots (in minutes), while the predicted synodic chopping
signal based on the published dynamical fit is shown in blue (also in
minutes).}
\label{fig:Kepler9_chopping0}
\end{figure}

The first system with detected transit timing variations is
the Kepler-9 system \citep{Kepler9}, which consists of
three planets, the outer two of which are close to 2:1 period
commensurability with periods near 19 and 39 days, respectively. The outer pair is dynamically decoupled from the inner planet in that the inner planet does not measurably affect their TTVs.
A recent dynamical analysis of Kepler-9 shows that the
transit timing variations yield masses of the outer two planets of 
$m_b = 45.1\pm1.5 M_\oplus$  and $m_c = 31.0\pm 1.0 M_\oplus$ \citep{Dreizler2014}.
We carried out a harmonic fit to the transit times for each planet, and find
an excellent fit to planet Kepler-9c[d]'s transit times with four[six] harmonics of
the period of Kepler-9d[c].  Figure \ref{fig:Kepler9_chopping0} shows the
transit timing variations for both planets along with the harmonic fit.
The synodic chopping ($j=1$) amplitude of the inner planet matches
the phase and amplitude predicted based on the mass inferred by \citet{Dreizler2014}.

We carried out a Markov chain analysis on the set of transit times
published by \citet{Dreizler2014} for this system with the 2:1 resonant term
and $j=1$ chopping, as well as additional harmonics with 
amplitudes that were not constrained by the physical parameters of
the model:  for the inner planet we added harmonics at $3n_2, 4n_2$,
and $6n_2$, while for the outer planet we added harmonics at $2n_1,
3n_1$, and $4n_1$ to our model.  We find similar masses as \citet{Dreizler2014}
$m_b = 49.9\pm 2.6 M_\oplus$ and $m_c = 35.6\pm 1.8 M_\oplus$, albeit with
larger uncertainties.  When we remove the constraint on the
amplitude of the chopping signal, we find comparable masses and 
uncertainties;  we suspect that the reason for this
is that in this case the amplitudes of the TTV are measured
with sufficient precision that the $f$ and $g$ terms
that occur in the first order resonant TTV formula can be
distinguished in amplitude.   The imaginary component of
the TTV is significant for both planets, so in this case
one can break the degeneracy between the planet masses
with just the resonant term.  The amplitudes for the
resonant term for both planets have four (well-measured)
constraints,  the real and imaginary amplitude for each
planet, while there are four unknowns: the mass ratios
for each planet, and the real and imaginary component
of $Z_{free}$.  This gives a unique solution, so the masses are well determined without the need
for the chopping constraint.

\begin{figure}
\center
\includegraphics[scale=0.75]{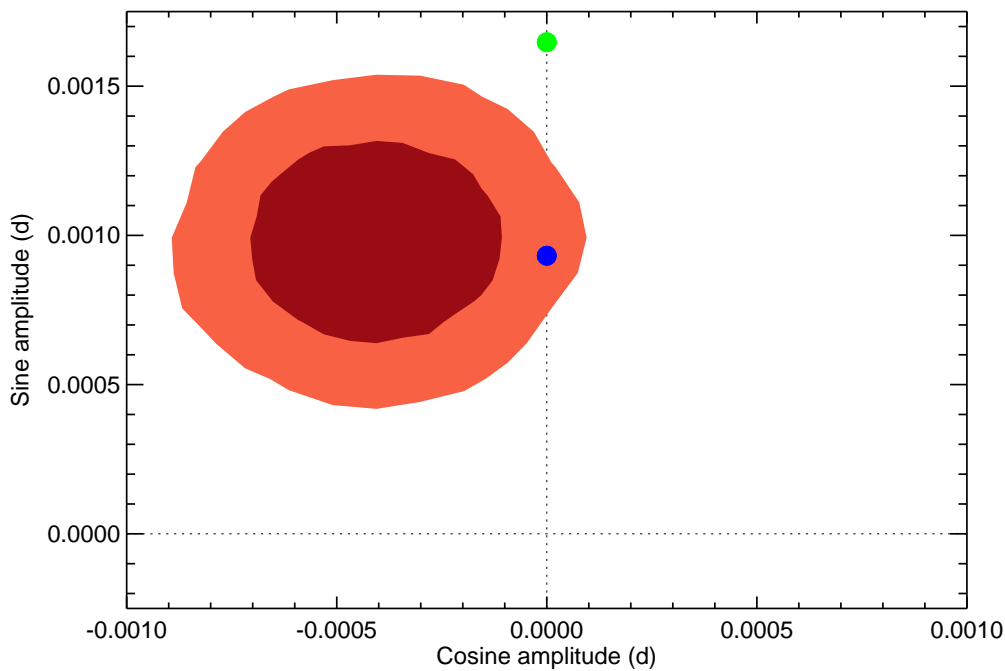}
\caption{Measured amplitude of the synodic chopping $\sin{(\lambda_1-\lambda_2)}$
term in the fit to Kepler-9, as well as the $\cos{(\lambda_1-\lambda_2)}$ term,
which is predicted to be zero; dark (light) red is 1(2)$\sigma$ confidence
region.  These are compared to the 2:1 resonant mass for $m_2 = 35.6 M_\oplus$ 
(blue; from TTVs) and $m_2 = 55 M_\oplus$ (green; from RV).}
\label{fig:Kepler9_chopping}
\end{figure}

Although chopping is not required to determine the mass
of the planets, we can show that in this case it is consistent
with the masses inferred from the resonant terms alone.  The measured mass of the outer planet predicts the
amplitude of the $j=1$ synodic chopping signal of the inner planet, which
only has $\sin{(\lambda_1-\lambda_2)}$ dependence.  In Figure \ref{fig:Kepler9_chopping}
we show that the measured sine amplitude of the inner chopping signal
is consistent to $<0.2\sigma$ with the predicted
amplitude.  The amplitude of the
synodic chopping term gives a mass of the outer planet
of $37.4\pm 8.6 M_\oplus$, while the mass estimated from
the resonant term is $35.6\pm 1.8 M_\oplus$.  In addition,
the amplitude of the cosine term is consistent with
zero at $<2\sigma$, as it should be for the synodic chopping term.

When radial velocities are included in the analysis, a larger mass 
is derived for the planets, yielding about 55 $M_\oplus$ for the
outer planet \citep{Dreizler2014}.  This is inconsistent with the chopping
signal, at about the $\approx$3$\sigma$ level (see green point in
Figure \ref{fig:Kepler9_chopping}) and inconsistent
with the masses derived from resonant TTV alone.  This
discrepancy indicates that there is still some tension between
the TTV data and the RV data, possibly due to RV
jitter, additional planets (causing perturbations of
the RV velocities), systematic
errors in the transit times (perhaps due to star spot
crossings), or, perhaps,  simply statistical fluctuations.  The
fact that the resonant and chopping terms give similar estimates
of the outer planet mass increases our confidence that the transit
timing analysis is not strongly affected by additional planets in
this system.  

For this system we also tried to use the amplitude of 
the $j=2$ chopping signal for the outer planet to constrain the mass 
of the inner planet.  However, the amplitude is {\it much} 
too large to be due to chopping, and instead we believe is 
due to the 2:4 resonant term (which is of order $e^2$, but 
these planets are so close to 1:2, that the $1/\Delta$ term 
compensates for this).

\subsection{Mass precision of KOI-872c}

For planets that are not near a first-order mean-motion
resonance ($P_1/P_2 \approx (j_R-1)/j_R$), the synodic
chopping amplitude can provide the strongest constraint
upon the planet masses.  If there are a large number
of transits, then the signal-to-noise of the planet
mass ratio can be estimated with equation (\ref{eqn:precision}).

\begin{figure}
\center
\includegraphics[scale=0.75]{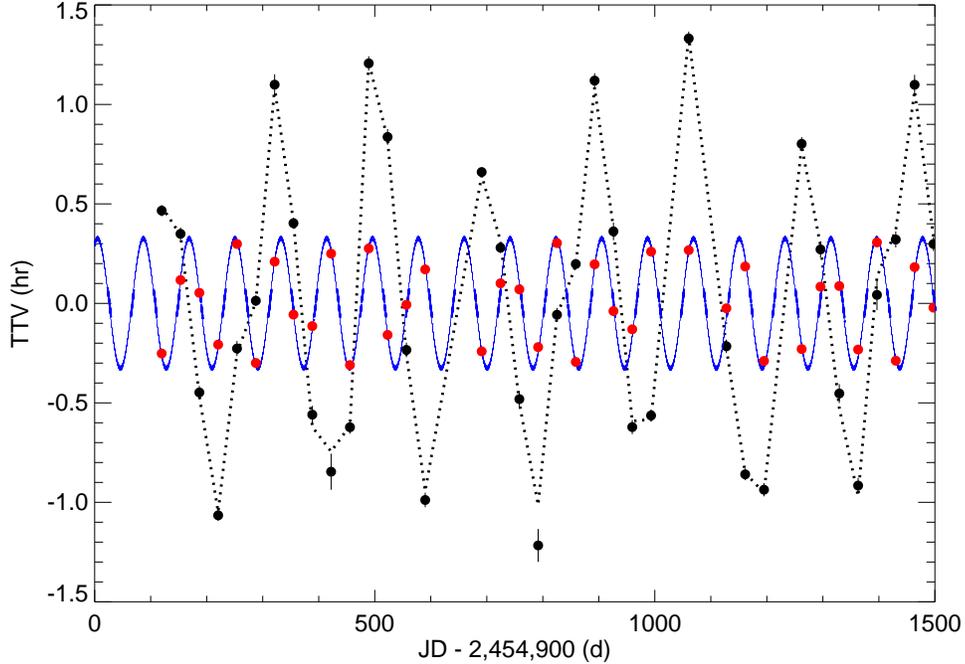}
\caption{Transit timing variations observed with Kepler for
KOI-872 (black filled circles, with uncertainties).  We have fit
five harmonics to the transit times, obtaining an excellent fit
(dotted line).  The synodic chopping component of our fit is
plotted with red dots, while the predicted synodic chopping
signal based on the published dynamical fit is shown in blue.
}
\label{fig:koi872_chopping}
\end{figure}

The first non-transiting planet found with transit
timing variations (with a unique identification of
the perturbing planet's period) occurred in the system
KOI-872 \citep{KOI872}, in which the period ratio of
the two planets is close to 5:3.  We carried out
a harmonic analysis of the transit times of the inner
planet with harmonics up to $5n_2$, starting with the
period of the perturbing planet, KOI-872c, from the published 
dynamical analysis.  The synodic chopping signal, the coefficient
of the $-n_2$ term, was measured to have an amplitude of
$a_1 = 0.0129\pm 0.0004$ days for the sinusoidal
component (detected at 32$\sigma$), and $b_1 =
-0.00049\pm0.00034$ for the cosine component (consistent
with zero at $\approx 1.5\sigma$).  Applying the
synodic chopping amplitude formula (Equation \ref{eqn:formulaInner},
with $j=1$), we find a mass ratio of: $m_2/M_\star= (3.81\pm 0.12) \times 10^{-4}$.
This compares favorably with the mass ratio measured from
the full dynamical analysis \citep{KOI872} of: $m_2/M_\star =
(3.97^{+0.15}_{-0.11}) \times 10^{-4}$, with a similar magnitude
uncertainty.  Figure \ref{fig:koi872_chopping} shows the
harmonic fit to the transit timing variations of KOI-872,
with the synodic chopping signal shown with red points.  The
predicted synodic chopping based upon the dynamical solution
is shown in blue, demonstrating agreement with
the derived signal (albeit discrepant by $\approx 1.4\sigma$).

The expected precision in the planet mass is given by equation \ref{eqn:precision};  for
KOI-872 there are 37 transit times with a typical precision of
0.0015 days, 13 model parameters, giving an expected mass precision 
of $\sigma_{M_2} = 1.25 \times 10^{-5}$, which matches well that found 
with the harmonic fits, and is close to the uncertainty found with
the full dynamical analysis.  This indicates that the timing
precision along with the total number of transits observed
and the number of free parameters fit can be used to forecast
the mass measurement precision in the non-resonant case.
A caveat is that this formula applies when chopping dominates
the mass uncertainty, which may not be true for large eccentricities
of the planets which can cause higher order resonant terms to
play a more important role.

\subsection{Predicting planetary mass precision inferred from the synodic TTV}
The chopping effect potentially allows for mass measurements of planets with TTVs observed by a TESS-like mission because the synodic TTV, when unaliased with a resonant frequency, is a short-period effect. However, the amplitude of the effect is also considerably smaller than that due to the long-timescale resonant variations, and so more transits are needed to build up signal to noise. Here we consider a system with 2 planets with a period ratio of 1.5, with an inner orbital period of 20 days. We assume that the system has been observed for 1 year (the baseline TESS will have for stars near the celestial poles), and assume that the timing uncertainty on the transit times of the inner planet are 1 minute.

In this case, $N_{trans} \approx 365.0(1/20.0+1/30.0) \approx 30 $ and the formula in Equation \ref{eqn:precision} yields, for 10 free parameters and assuming a solar mass star, a mass precision of the outer planet based on the $j=1$ chopping of the inner planet of $\sim 1.3 M_\oplus$. The mass precision on the inner planet due to the $j=1$ chopping in the outer planet is $\sim 1 M_\oplus$. Wider pairs will have larger mass uncertainty since the function $f^j_i(\alpha)$ is smaller. For example, this same pair moved to the 2:1 resonance will allow an $\sim 6M_\oplus$ mass uncertainty for the outer planet.

 Note that the mass uncertainty scales like the inverse of the square of the orbital period, and, given an orbital period, like the inverse of the square of the observation time. Therefore, for a longer mission like PLATO, with an observational baseline of 2 or 3 years, the uncertainty on the mass of the perturbing planet will be smaller by a factor of $1/\sqrt{2}$ or $1/\sqrt{3}$, respectively, compared with that estimated for 1 year of data above.

\section{Conclusions}\label{sec:conclusion}

In this paper we have written down expressions
for transit-timing variations in the plane-parallel
limit, in the limit of zero free eccentricity of both 
planets, outside of resonance, and for timescales
shorter than the secular timescale.  Despite these assumptions, these terms have important
consequences for analysis of transit-timing variations
of multi-planet systems:  1) the TTVs have a 
dependence on $\sin{\left[j(\lambda_1-\lambda_2)\right]}$,
for $j=1$ to $\infty$; 2) the amplitude of these
terms only depends on the mass-ratio of the perturbing
planet to the star, and the semi-major axis ratio of
these planets.  Although other papers have presented
formulae in the limit of zero-eccentricity \citep{Agol2005,
Nesvorny2014}, this is the first time that the coefficients 
for each $j$ have been written down explicitly.  This
allows for harmonic analysis of transit times in terms
of the period and phase of the perturbing planet; the coefficients
of each harmonic can then be related to the planet mass
ratios and eccentricities via these formulae (except for
the harmonics affected by resonant terms).  When
the period and phase of the perturbing planet are known,
then fitting for the harmonic coefficients (and
the mean ephemeris) is a linear regression problem;  
thus a global solution can be found by simple 
matrix inversion, yielding a unique solution for the coefficients.
This means that there cannot be multi-modal degeneracies
in the derived masses of the planets.
The amplitudes and uncertainties of these coefficients can
then be translated into planet masses using the formulae
given above.  Alternatively the coefficients can computed
from the physical properties of the planets (masses, eccentricity vector,
and ephemerides), and then the model can be fit to the data with
a non-linear optimization or Markov chain, as we have carried
out for several of the examples presented above.

In particular, we have highlighted the importance
of the $j=1$ `synodic chopping' signal of the inner planet, 
which has a weak dependence on the eccentricities of the planets,
and is not aliased with any first-order resonant terms. The formula for this term in particular has very little error even for very compact orbits.  For the outer
planet, the $j=1$ chopping signal can be used if the
period ratios are distant from the 2:1 resonance, though if the transit times are precisely measured the synodic TTV resulting from values of $j \neq 1$ may be used to constrain the mass of the inner planet instead. In general, if the system is near or in a first order resonance, with $P_2/P_1 \approx (j_R-1)/j_R$, only the synodic TTV component with $j=j_R$ (inner planet) and $j=j_R-1$ (outer planet) will be altered; the formulae with $j$ not equal to these values will apply.  Other
limitations of this formula is that it cannot be directly used
when more than one planet strongly perturbs the transiting
planet, it breaks down for very massive planets ($m_{planet}\sim 10^{-3} M_\star$) with period ratios less than 2, and that as the period ratio approaches
unity the error due to neglected eccentricity terms can become more important in the formula (though this does not apply to the $j=1$ term in the TTVs of the inner planet). Following \citet{Nesvorny2014}, we conclude that in these regimes the chopping formula provides insight into mass measurements with TTVs even though the analytic formulae may be too approximate to be applied directly.

We have applied these formulae to existing transiting
planet systems that have been analyzed in prior publications,
recovering the mass measurements, but using harmonic fits
and the analytic chopping/resonant expressions rather
than full N-body integrations.  In some cases (KOI-872) this shows that the primary constraint on
the mass of the planets comes from the synodic chopping
signal.  In other cases (KOI-1353c/d, Kepler 11d,e) the
primary mass constraint comes from the combination of the first-order
resonant signal and the synodic chopping of the inner planet.
In the case of Kepler-9, the primary constraint comes
from the resonant signal, although the chopping component
gives a consistent constraint on the mass of the outer
planet.

In future applications, we expect that these formulae
can be used for rapid fitting and estimation of transit timing variations,
for rapid estimation of planet masses, for initialization
of N-body integration, for determining the requisite
timing precision to measure planet masses with future
follow-up observations of multi-planet transiting systems,
and for forecast of transit timing variations. It should
be possible to apply these formulae in systems of more
than two planets using linear-combinations of the TTVs
induced by more than one perturbing planet.

\acknowledgments
We would like to thank Dan Fabrycky, Eric Ford, Matt Holman, Daniel Jontof-Hunter, Jack Lissauer, Jason Steffen, and the {\it Kepler} TTV group for helpful conservations.  E.A. acknowledges funding by NSF Career
Grant AST 0645416, NASA Astrobiology Institute's Virtual
Planetary Laboratory, supported by NASA under cooperative
agreement NNH05ZDA001C, and NASA Origins of Solar Systems 
Grant 12-OSS12-0011.  K.M.D. acknowledges supports from JCPA fellowship at Caltech. Work by K.M.D was supported by NASA under grant NNX09AB28G from the Kepler Participating Scientist Program and grants
NNX09AB33G and NNX13A124G under the Origins program.

\bibliographystyle{apj}

\bibliography{submittedDeck_Agol.bib}

\appendix
\section{Derivation of synodic TTV}
Here we provide an alternate derivation of the synodic TTV to zeroth order in planetary eccentricities. We follow the method first described in \citet{NesvornyMorbidelli2008}, and work at zeroth order in eccentricities and inclinations and first order in the parameter $m_{planet}/M_\star$.

As mentioned in the main text, the synodic TTV has been derived before, by \citet{Agol2005} and \citet{Nesvorny2014}. These, and the derivation below, give agreement 
in the limit that the reflex motion of the star can be ignored
(after correcting a typo in \citet{Nesvorny2014}).  For the outer planet, the reflex
motion of the star dominates at large period ratios.  This effect is accounted for
in the equations in the appendix of \citet{Agol2005} and in the alternate derivation in the appendix here, but is missing from the
terms in \citet{Nesvorny2014} (the \citet{Agol2005} calculation utilized heliocentric coordinates,
while that of \citet{Nesvorny2014} used Jacobi coordinates and thus computed the TTVs of the outer
planet relative to the center of mass, not relative to the star. Here we also employ Jacobi coordinates but correct for the reflex contribution afterwards.

TTVs are determined by taking an observed set of transit times and performing a linear fit to them, such that the transit timing variation $\delta_t$ can be written as
\begin{align}
\delta t_n & = t_n-(t_0+n \bar{P})
\end{align}
The slope $\bar{P}$ is the average time between successive transits during the observational baseline, or the average period over this timespan. In this sense, transit timing variations are the deviations from the transit times predicted by the average of the true (non-Keplerian) orbit.  

Transits occur when the true longitude $\theta$ of the transiting planet is equal to a particular value. The expression for $\theta$ is, to first order in eccentricity,
\begin{align}\label{Eqn1}
\theta & = f+\varpi \nonumber \\
& = M+2 e \sin{M}+\varpi + O(e^2) \nonumber \\
& = \lambda + 2 e \sin{(\lambda-\varpi)}+O(e^2)
\end{align}
where $e$ is the eccentricity, $f$ is the true anomaly, $\varpi$ the longitude of periastron, $M$ the mean anomaly, and  $\lambda=M+\varpi$ the mean longitude of the planet. Because we will perform the calculation using a Hamiltonian formalism, we now switch to canonical coordinates. Our variables will be
\begin{center}\label{variables}
\begin{tabular}{l c l}
$\Lambda_i  = m_i\sqrt{G M_\star a_i}$ &      &$ \lambda_i$  \\
$x_i  = \sqrt{2P_i}\cos{p_i} $&     &$ y_i =\sqrt{2P_i}\sin{p_i}$ \\
\end{tabular}
\end{center}
where:
\begin{center}\label{variables2}
\begin{tabular}{lcr}
$P_i = \Lambda_i \frac{e_i^2}{2} + O(e_i^4) $&     & $p_i = -\varpi_i$
\end{tabular}
\end{center}
where $i=0,1,$ for the two planets, $m_i$ is the mass of the planet, $M_\star$ the mass of the star, and all orbital elements are Jacobi elements. Variables in the left columns are the canonical momenta, while those in the right columns are the conjugate coordinate.
In terms of this canonical set, Equation \ref{Eqn1} for $\theta$ becomes
\begin{align}\label{Eqn2}
\theta & = \lambda + \frac{2}{\sqrt{\Lambda}}\bigg( x \sin{\lambda}+y \cos{\lambda} \bigg)
\end{align}

We will perturb Equation \ref{Eqn2} about the averaged orbit, and keep terms only at zeroth order in $x$ and $y$:
\begin{align}\label{Eqn3}
\delta \theta & = \delta \lambda + \frac{2}{\sqrt{\Lambda}}\bigg( \delta x \sin{\lambda}+\delta y \cos{\lambda}  \bigg)
\end{align}
where as we will see the parameters $\delta$ are going to be of order $\epsilon = m_{planet}/M_\star$ and independent of eccentricities to lowest order.  It is important to note that we define $\delta f \equiv f-\bar{f}$ for any arbitrary function $f$, where over-bar denotes a time-average. In that case, 
\begin{align}
\delta \theta & = -n \delta t
\end{align}
where $\bar{n}$ is the average mean motion and $\delta t$ is the timing variation. Note that 1) we are consistent with the style that a negative timing variation corresponds to an ``early" transit with $\delta \theta >0$ and 2) we can use the average mean motion because Equation \ref{Eqn3} holds only at first order in the perturbations; in the same sense, we can treat quantities without a $\delta$ on the right hand side as averaged variables as well.

The question now arises - how do we determine the perturbations to the average orbit $\delta \lambda, \delta x$ and $\delta y$? 

The Hamiltonian for a system of two planets of mass $m_i$ orbiting a much more massive star of mass $M_\star$, written in Jacobi coordinates $\mathbf{{r}}_i$ and momenta $\mathbf{{p}}_i$, takes the form, to first order in combinations of $\epsilon = m_{planet}/M_\star$, of
\begin{align}
{H} & = {H}_0 + {H}_1 \nonumber \\
{H}_0 & = {H}_{Kepler ,1} + {H}_{Kepler, 2} \nonumber \\
{H}_{Kepler, i} & =  \frac{{p}_i^2}{2\tilde{m}_i}   -\frac{G \tilde{M}_{i,\star} \tilde{m}_i}{\vert \mathbf{{r}}_i \vert} \nonumber \\
{H}_1 & = -G {m}_1 {m}_2\left(\frac{1}{\vert  \mathbf{{r}}_1 - \mathbf{{r}}_2 \vert}  - \frac{ \mathbf{{r}_2} \cdot \mathbf{{r}}_1}{\vert \mathbf{{r}}_2\vert^3} \right) +O(\epsilon^2),
\end{align}
where $\tilde{M}_{2,\star} = M_\star(M_\star+m_1+m_2)/(M_\star+m_1) $, $\tilde{M}_{1,\star} =  (M_\star+m_1) $ and $\tilde{m}_i =m_i+O(\epsilon)$ denote Jacobi masses. Note that the perturbation $H_1$ takes the functional form of the disturbing function with an exterior perturber \citep{MurrayDermott}. We set $\tilde{M}_{i,\star}= M_\star$ and also ignore the difference between Jacobi and physical masses.  In the Keplerian piece ${H}_0$, this approximation corresponds to a (constant) change in the mean motions of the planets by order $\epsilon$, but since we are interested in TTVs this constant change does not matter. The correction between Jacobi and physical masses in the perturbation $H_1$ generates only $O(\epsilon^2)$ terms, and we ignore these. 

 When expressed in terms of the canonical set given in Equation \ref{variables} and Equation \ref{variables2}, the Hamiltonian takes the form:
\begin{align}
H & = H_0({\Lambda}_i) + \epsilon_1 H_1({\Lambda}_i,{P}_i,{\lambda}_i,{p}_i) + O(\epsilon^2)
\end{align}
where e.g. ${\Lambda}_i =(\Lambda_1,\Lambda_2)$, $\epsilon_1 = \frac{m_1}{M}$, and
\begin{align}\label{FullA}
 H_0 & = -\frac{\mu_1}{2\Lambda_1^2} -\frac{\mu_2}{2\Lambda_2^2} \nonumber \\
 H_1 & = -\frac{\mu_2}{\Lambda_2^2}\bigg[ \sum_{j=-\infty}^{j=\infty} g_{j,0}(\alpha) \cos{j(\lambda_1-\lambda_2)} +\nonumber \\
 & \sum_{j=-\infty}^{j=\infty} g_{j,27}(\alpha)\sqrt{\frac{2P_1}{\Lambda_1}} \cos{(j\lambda_2 -(j-1)\lambda_1+p_1)} +\nonumber\\
 &\sum_{j=-\infty}^{j=\infty} g_{j,31}(\alpha)\sqrt{\frac{2P_2}{\Lambda_2}} \cos{(j\lambda_2 -(j-1)\lambda_1+p_2)}  \bigg]
 \end{align}
 and  $\mu_i = G^2 M_\star^2 m_i^3$, $\alpha = a_1/a_2$, $\sqrt{2P/\Lambda} = e(1+O(e^2))$, and $g_{j,27}$, $g_{j,31}$ and $g_{j,0}$ are functions of Laplace coefficients with the indirect terms included \citep{MurrayDermott}\footnote{We could equivalently use $\epsilon_2$ instead of $\epsilon_1$ as our small parameter, the overall coefficient of $H_1$ would just change to $-\bigg(m_2/m_1 \bigg)^2\mu_1/\Lambda_2^2$.}.  The relevant functions of Laplace coefficients can be written as
 \begin{align}
 g_{j,0}(\alpha) & = \frac{1}{2}b_{1/2}^j(\alpha)- \delta_{j,1} \alpha \nonumber \\ 
 g_{j,27}(\alpha) &= \frac{1}{2}(-2 j-\alpha \frac{d}{d\alpha}) b_{1/2}^j(\alpha) + \delta_{j,1} \frac{3}{2}\alpha - \delta_{j,-1} \frac{1}{2}\alpha\nonumber \\
 g_{j,31}(\alpha) &= \frac{1}{2}(-1+2j+\alpha \frac{d}{d\alpha}) b_{1/2}^{j-1}(\alpha)-\delta_{j,2} 2\alpha \nonumber \\
 b_{1/2}^j(\alpha) & = \frac{1}{\pi} \int_0^{2\pi} \frac{\cos{(j \theta)}}{\sqrt{1-2\alpha \cos{\theta}+\alpha^2}}d \theta.
 \end{align} 

In writing the gravitational potential in this form, we have assumed that $\alpha$ is not too close to unity, in which case Laplace coefficients converge slowly. For $\alpha \rightarrow 1$, the analogs of $g_{j,0},g_{j,27},$ and $g_{j,31}$ appearing with higher powers of eccentricity are larger and so the neglected terms in Equation \ref{FullA} become more important. 

   We now convert the $(P_i,p_i)$ variables (Equation \ref{variables2}) to the $(x_i,y_i)$ set (Equation \ref{variables}), and also set $j(\lambda_1-\lambda_2) = \psi_j$ and $j\lambda_2 -(j-1)\lambda_1 = \phi_j$. Then the perturbation Hamiltonian $H_1$ takes the form:
  \begin{align}\label{Full}
  H_1  = -\frac{\mu_2}{\Lambda_2^2}\bigg[ \sum_{j=-\infty}^{j=\infty} g_{j,0}(\alpha) \cos{\psi_j} &+  \sum_{j=-\infty}^{j=\infty} g_{j,27}(\alpha)\frac{1}{\sqrt{\Lambda_1}} (x_1 \cos{\phi_j}-y_1 \sin{\phi_j}) \nonumber\\
 &+ \sum_{j=-\infty}^{j=\infty} g_{j,31}(\alpha)\frac{1}{\sqrt{\Lambda_2}} (x_2 \cos{\phi_j}-y_2 \sin{\phi_j})  \bigg]
 \end{align}

 If the system is not too close to any resonance, all periodic terms in the perturbation $H_1$ are short-period, and an average over the (fast) angles $\lambda_1-\lambda_2$ and e.g. $\lambda_2$ leads to the averaged Hamiltonian $\bar{H}$ which is only a function of $(\bar{\Lambda}_i)$ (the assumption that all terms are short-period will be discussed at the end of this section). This averaged Hamiltonian takes the same function form of $H_0(\bar{\Lambda}_i)$, and so the evolution of the averaged orbits is ``Keplerian": only the $\bar{\lambda}_i$ vary, and they do so at a constant rate given by $\bar{n}_i = \sqrt{GM_\star/\bar{a}_i^{3}}$). These are the averaged orbits we are interested in for calculating TTVs, and hence we need to determine a canonical transformation which turns $H$ into $\bar{H}$ and hence also changes exact variables into averaged ones. 
 
 This canonical transformation requires removing the short period terms via a generating function so that they do not appear in the averaged Hamiltonian.  The deviation between the averaged and full trajectories is of order $\epsilon$, and hence the canonical transformation between the two sets of variables should be very near the identity transformation. Therefore, we seek a generating function of Type 2 which determines a near-identity transformation that removes the short period terms. We will write this generating function as
\begin{align}
F_2(\lambda_i,y_i,\bar{\Lambda}_i,\bar{x}_i) &= \lambda_i \bar{\Lambda}_i + y_i\bar{x}_i + \epsilon_1 \it{f}(\lambda_i,y_i,\bar{\Lambda}_i,\bar{x}_i) 
\end{align}
where $f$ is yet to be determined.

Again the variables with over-bars are the averaged canonical coordinates and momenta, while those without correspond to the unaveraged variables. The old and new variables are related as
\begin{align}\label{transform}
\Lambda_i & = \frac{\partial F_2}{\partial \lambda_i} = \bar{\Lambda}_i+\epsilon_1\frac{\partial \it{f}}{\partial \lambda_i} \nonumber \\
x_i & = \frac{\partial F_2}{\partial y_i} = \bar{x}_i+\epsilon_1\frac{\partial \it{f}}{\partial y_i} \nonumber \\
\bar{\lambda}_i & = \frac{\partial F_2}{\partial \bar{\Lambda}_i} = \lambda_i+\epsilon_1\frac{\partial \it{f}}{\partial \bar{\Lambda}_i} \nonumber \\
\bar{y}_i & = \frac{\partial F_2}{\partial \bar{x}_i} = y_i+\epsilon_1\frac{\partial \it{f}}{\partial \bar{x}_i} 
\end{align}
so that the differences we are seeking for Equation \ref{Eqn3} are
\begin{align}
\delta \lambda_i & = \lambda_i-\bar{\lambda}_i  = -\epsilon_1\frac{\partial \it{f}}{\partial \bar{\Lambda}_i} \nonumber \\
\delta x_i & = x_i-\bar{x}_i  = \epsilon_1\frac{\partial \it{f}}{\partial y_i} \nonumber \\
\delta y_i & = y_i-\bar{y}_i  = -\epsilon_1\frac{\partial \it{f}}{\partial \bar{x}_i}
\end{align}
Then, the expression for $\delta \theta_i$ is simply:
\begin{align}\label{bigeqn}
\delta \theta_i  = \epsilon_1\bigg[&-\frac{\partial \it{f}}{\partial \bar{\Lambda}_i}+ \frac{2}{\sqrt{\Lambda_i}}\bigg(\frac{\partial \it{f}}{\partial y_i}\sin{\lambda_i}-\frac{\partial \it{f}}{\partial \bar{x}_i} \cos{\lambda_i}  \bigg)\bigg]
\end{align}
where we can evaluate the function $\it{f}(\lambda_i,y_i,\bar{\Lambda}_i,\bar{x}_i)$ as $\it{f}(\bar{\lambda}_i,\bar{y}_i,{\bar{\Lambda}}_i,\bar{x}_i)$ because the difference between the averaged and unaveraged variables is already of order $\epsilon$.

The problem now reduces to finding the function $f$.  The new (averaged) Hamiltonian is $\bar{H} (\bar{\Lambda}_i,\bar{x}_i,\bar{\lambda}_i,\bar{y}_i) = H[\bar{\Lambda}_i({\Lambda}_i,{x}_i,{\lambda}_i,{y}_i),\bar{x}_i({\Lambda}_i,{x}_i,{\lambda}_i,{y}_i),\bar{\lambda}_i({\Lambda}_i,{x}_i,{\lambda}_i,{y}_i),\bar{y}_i({\Lambda}_i,{x}_i,{\lambda}_i,{y}_i))]$ since the generating function is time independent. To first order in $\epsilon_1$,
\begin{align}
\bar{H} (\bar{\Lambda}_i,\bar{x}_i,\bar{\lambda}_i,\bar{y}_i) = H_0\bigg(\bar{\Lambda}_1+&\epsilon_1\frac{\partial \it{f}}{\partial \lambda_1},\bar{\Lambda}_2+\epsilon_1\frac{\partial \it{f}}{\partial \lambda_2}\bigg) +    \epsilon_1 H_1(\bar{\Lambda}_i,\bar{x}_i,\bar{\lambda}_i,\bar{y}_i) \nonumber \\
=H_0(\bar{\Lambda}_i)+ & \epsilon_1 \bigg( \frac{\partial H_0}{\partial \Lambda_1} \frac{\partial \it{f}}{\partial \lambda_1} +  \frac{\partial H_0}{\partial \Lambda_2} \frac{\partial \it{f}}{\partial \lambda_2} \bigg)\bigg \lvert_{(\Lambda_i,x_i, \lambda_i,y_i) = (\bar{\Lambda}_i, \bar{x}_i,\bar{\lambda}_i,\bar{y}_i)} +\epsilon_1H_1(\bar{\Lambda}_i, \bar{x}_i,\bar{\lambda}_i,\bar{y}_i)\nonumber \\
  =  H_0(\bar{\Lambda}_i) +& \epsilon_1 (\bar{n}_{1} \frac{\partial \it{f}}{\partial \lambda_1} +\bar{n}_{2} \frac{\partial \it{f}}{\partial \lambda_2}) \bigg \lvert_{(\Lambda_i,x_i, \lambda_i,y_i) = (\bar{\Lambda}_i, \bar{x}_i,\bar{\lambda}_i,\bar{y}_i)} +\epsilon_1H_1(\bar{\Lambda}_i, \bar{x}_i,\bar{\lambda}_i,\bar{y}_i)
 \end{align}
where $\bar{n}_{i} = \mu_i^2/\bar{\Lambda}_i^3$ is the average mean motion of planet $i$.

Now, we would like this averaged Hamiltonian to be equal to
\begin{align}
\bar{H} & =   H_0(\bar{\Lambda}_i)
\end{align}
i.e. with all short period terms removed. We must choose the function $f$ to satisfy the relation:
\begin{align}\label{satisfy}
 (\bar{n}_{1} \frac{\partial \it{f}}{\partial \lambda_1} &+\bar{n}_{2} \frac{\partial \it{f}}{\partial \lambda_2})  + H_1 =0
\end{align}

Given an $H_1$ that takes the functional form $H_1 = A \cos{\theta}+ B \sin{\theta}$, where $A$ and $B$ are coefficients independent of $\lambda_i$, and $\theta = p\lambda_1 + q \lambda_2$, for any integers $p$ and $q$, the solution $f$ to Equation \ref{satisfy} is
\begin{align}
f = \frac{1}{\dot{\theta}}\bigg[ -A \sin{\theta}+ B \cos{\theta}\bigg]
\end{align}
where $\dot{\theta} = p \bar{n}_1+q \bar{n}_2$.  
Therefore, the solution to Equation \ref{satisfy} for the full $H_1$ given in Equation \ref{Full} is:
\begin{align}
f &=  -\frac{\mu_2}{\bar{\Lambda}_2^2}  \sum_{j} -\bigg(\frac{g_{j,0}(\bar{\alpha})\sin{\psi_j}}{j(\bar{n}_1-\bar{n}_2)} +\frac{g_{j,27}(\bar{\alpha})}{\sqrt{\bar{\Lambda}_1}(j \bar{n}_2 -(j-1) \bar{n}_1)} (\sin{\phi_j} \bar{x}_1 +\cos{\phi_j} y_1) \nonumber \\
&+\frac{g_{j,31}(\bar{\alpha})}{\sqrt{\bar{\Lambda}_2}(j \bar{n}_2 -(j-1) \bar{n}_1)} (\sin{\phi_j} \bar{x}_2 +\cos{\phi_j} y_2) \bigg)
\end{align}
Note that the first term of the three diverges when $j=0$ (in the Hamiltonian, this is the only pieces of the interaction term which is independent of the orbital angles). We will ignore this term, since it only leads to a constant change in the mean motion of the planets. At this point, we will also stop denoting averaged variables with over-bars, since at the order we are working in $\epsilon$ these two are equivalent. 
Then, 
\begin{align}\label{derivofF}
\frac{\partial \it{f}}{\partial \bar{\Lambda}_1}   =   \frac{\mu_2}{{\Lambda}_2^2} & \sum_{j} \bigg( \frac{d g_{j,0}(\alpha)}{d \alpha} \frac{\partial \alpha/\partial \Lambda_1}{j(n_1-n_2)} - g_{j,0}(\alpha) \frac{\partial n_1/\partial \Lambda_1}{j(n_1-n_2)^2}  \bigg)\sin{\psi_j}  +O(e) \nonumber \\
 = \frac{\mu_2}{{\Lambda}_2^2}  &\sum_{j} \bigg( \frac{d g_{j,0}(\alpha)}{d \alpha} \frac{2\alpha/\Lambda_1}{j(n_1-n_2)} - g_{j,0}(\alpha) \frac{-3n_1/\Lambda_1}{j(n_1-n_2)^2}  \bigg)\sin{\psi_j}  +O(e) \nonumber \\
 = \frac{n_2}{n_1} \frac{\Lambda_2}{\Lambda_1}&\sum_{j}  \bigg( \frac{d g_{j,0}(\alpha)}{d \alpha} \frac{2\alpha n_1}{j(n_1-n_2)} + g_{j,0}(\alpha) \frac{3n_1^2}{j(n_1-n_2)^2}  \bigg)\sin{\psi_j}  +O(e) \nonumber \\
 = \frac{m_2}{m_1} \alpha &\sum_{j}  \bigg( \frac{d g_{j,0}(\alpha)}{d \alpha} \frac{2 \alpha}{ \beta} + g_{j,0}(\alpha) \frac{3 j}{ \beta^2}  \bigg)\sin{\psi_j}  +O(e)
\end{align}
where $\beta = j(n_1-n_2)/n_1$, and
\begin{align}
\frac{\partial \it{f}}{\partial \bar{\Lambda}_2} = \frac{\mu_2}{{\Lambda}_2^2}  &\sum_{j} \bigg( \frac{d g_{j,0}(\alpha)}{d \alpha} \frac{\partial \alpha/\partial \Lambda_2}{j(n_1-n_2)} + g_{j,0}(\alpha) \frac{\partial n_2/\partial \Lambda_2}{j(n_1-n_2)^2}  \bigg)\sin{\psi_j} \nonumber \\
- 2 n_2 & \sum_{j} \frac{g_{j,0}(\alpha)}{j(n_1-n_2)}\sin{\psi_j}+ O(e) \nonumber \\
 = n_2 &\sum_{j} \bigg( \frac{d g_{j,0}(\alpha)}{d \alpha} \frac{-2 \alpha }{j(n_1-n_2)}  - g_{j,0}(\alpha) \frac{3 n_2}{j(n_1-n_2)^2} -2 \frac{g_{j,0}(\alpha)}{j(n_1-n_2)} \bigg)\sin{\psi_j} + O(e)  \nonumber \\
 = -&\sum_{j} \bigg( \frac{d g_{j,0}(\alpha)}{d \alpha} \frac{2 \alpha }{\kappa}  +g_{j,0}(\alpha) \frac{3 j}{\kappa^2} +2 \frac{g_{j,0}(\alpha)}{\kappa} \bigg)\sin{\psi_j}  +O(e)
\end{align}
where $\kappa = j(n_1-n_2)/n_2$.

We now calculate the changes in $x_i$ and $y_i$ (again ignoring the difference between averaged and unaveraged variables at this order in $\epsilon$):
\begin{align}\label{derivofF2}
\frac{\partial \it{f}}{\partial y_1} & = \frac{\mu_2}{{\Lambda}_2^2}  \sum_{j} \frac{g_{j,27}(\alpha)}{\sqrt{{\Lambda}_1}(j n_2 -(j-1) n_1)} \cos{\phi_j} \nonumber \\
\frac{\partial \it{f}}{\partial y_2} & = \frac{\mu_2}{{\Lambda}_2^2}  \sum_{j} \frac{g_{j,31}(\alpha)}{\sqrt{{\Lambda}_2}(j n_2 -(j-1) n_1)} \cos{\phi_j} \nonumber \\
\frac{\partial \it{f}}{\partial x_1} & = \frac{\mu_2}{{\Lambda}_2^2}  \sum_{j} \frac{g_{j,27}(\alpha)}{\sqrt{{\Lambda}_1}(j n_2 -(j-1) n_1)} \sin{\phi_j} \nonumber \\
\frac{\partial \it{f}}{\partial x_2} & = \frac{\mu_2}{{\Lambda}_2^2}  \sum_{j} \frac{g_{j,31}(\alpha)}{\sqrt{{\Lambda}_2}(j n_2 -(j-1) n_1)} \sin{\phi_j}
\end{align}

Ultimately, we want the combinations:
\begin{align}
Z_i & = \frac{2}{\sqrt{\Lambda_i}}\bigg( \delta x_i  \sin{\lambda_i}-\delta y_i \cos{\lambda_i}  \bigg) \nonumber \\
& = \frac{2 \epsilon_1}{\sqrt{\Lambda_i}} \bigg(\frac{\partial \it{f}}{\partial y_i} \sin{\lambda_i}-\frac{\partial \it{f}}{\partial x_i} \cos{\lambda_i}  \bigg)
\end{align}

For the inner planet, this  is
\begin{align}
Z_1 & =  \epsilon_1 \frac{\mu_2}{{\Lambda}_2^2} \frac{2}{\Lambda_1}\sum_j \frac{g_{j,27}(\alpha)}{j n_2 -(j-1) n_1}\bigg(\cos{\phi_j} \sin{\lambda_1} - \sin{\phi_j} \cos{\lambda_1 }\bigg)  \nonumber \\
& = 2\epsilon_1 n_2 \frac{\Lambda_2}{\Lambda_1} \sum_j \frac{g_{j,27}(\alpha)}{j n_2 -(j-1) n_1}\sin{(\lambda_1-\theta_j)} \nonumber \\
& = 2\epsilon_1 \frac{n_2}{n_1} \frac{m_2}{m_1}\alpha^{-1/2}\sum_j \frac{g_{j,27}(\alpha) n_1 }{j n_2 -(j-1) n_1}\sin{\psi_j}\nonumber \\
& = 2\frac{m_2}{m_1}\epsilon_1 \alpha \sum_j \frac{g_{j,27}(\alpha)}{1-\beta}\sin{\psi_j}
\end{align}
Note that only a synodic dependence of $\psi_j$ remains! At zeroth order in eccentricity, there are no first-order (and possibly resonant) angles $\phi_j$, although we needed these terms to compute the final result! The final expression for $\delta \theta_1$ is given by:
\begin{align}
-n_1 \delta t_1 =\delta \theta_1 & = Z_1 + \delta \lambda_1 \nonumber \\
 \delta \theta_1&= \frac{m_2}{m_1}\alpha \epsilon_1 \sum_j \bigg(2\frac{g_{j,27}(\alpha)}{1-\beta} -\frac{d g_{j,0}(\alpha)}{d \alpha} \frac{2 \alpha}{ \beta} - g_{j,0}(\alpha) \frac{3 j}{ \beta^2} \bigg)\sin{\psi_j} \nonumber \\
\delta t_1 & = \frac{P_1}{2\pi} \epsilon_2 \alpha  \sum_{j=-\infty}^{j=\infty} \bigg(-2\frac{g_{j,27}(\alpha)}{1-\beta} +\frac{d g_{j,0}(\alpha)}{d \alpha} \frac{2 \alpha}{ \beta} +g_{j,0}(\alpha) \frac{3 j}{ \beta^2} \bigg)\sin{\psi_j} \nonumber \\
\delta t_1 &= \frac{P_1}{2\pi} \epsilon_2 \alpha  \sum_{j=1}^{\infty} \bigg( \bigg[\frac{d g_{j,0}(\alpha)}{d \alpha}+\frac{d g_{-j,0}(\alpha)}{d \alpha} \bigg] \frac{2 \alpha}{ \beta}+\bigg[g_{j,0}(\alpha)+g_{-j,0}(\alpha)\bigg] \frac{3 j}{ \beta^2} +\nonumber \\
& 2\bigg[-\frac{g_{j,27}(\alpha)}{1-\beta}+ \frac{g_{-j,27}(\alpha)}{1+\beta}\bigg]\bigg)\sin{\psi_j} \nonumber \\
\delta t_1 &= \frac{P_1}{2\pi} \epsilon_2 \alpha  \sum_{j=1}^{\infty}\bigg(  \frac{2}{\beta} \bigg[D_\alpha b^j_{1/2}(\alpha) - \alpha \delta_{j,1}\bigg] + \frac{3j}{\beta^2} \bigg[ b^j_{1/2}(\alpha) - \alpha \delta_{j,1}\bigg] \nonumber \\
&+ \frac{2}{(1-\beta^2)} \bigg[ (2j b^j_{1/2}(\alpha)+\beta D_\alpha b^j_{1/2}(\alpha)-\alpha \delta_{j,1}(2+\beta)\bigg]\bigg) \nonumber \\
\delta t_1 & = \frac{P_1}{2\pi} \epsilon_2 \alpha \frac{2\beta D_\alpha  b^j_{1/2}(\alpha) + j(3+\beta^2)  b^j_{1/2}(\alpha)-\alpha \delta_{j,1}(\beta^2+2\beta+3) }{\beta^2 (1-\beta^2)}
\end{align}
where we have neglected the constant $j=0$ contribution, defined $D_\alpha(\cdot) = \alpha \partial (\cdot)/\partial \alpha$, and taken advantage of the fact that as $j\rightarrow -j$, $\beta$ changes sign and the Laplace coefficients $b^j_{1/2}$ and their derivatives do not.
Therefore, 
\begin{align}\label{f1}
\delta t_1 & = \frac{P_1}{2\pi} \epsilon_2 \sum_{j=1}^{\infty} f_1^j(\alpha))\sin{\psi_j}  \nonumber \\
f_1^j(\alpha) & = \alpha \frac{2\beta D_\alpha  b^j_{1/2}(\alpha) + j(3+\beta^2)  b^j_{1/2}(\alpha)-\alpha \delta_{j,1}(\beta^2+2\beta+3) }{\beta^2 (1-\beta^2)}
\end{align}

For the outer planet, the combination $Z_2$ is:
\begin{align}
Z_2  =  \epsilon_1 \frac{\mu_2}{{\Lambda}_2^2} \frac{2}{\Lambda_2}&\sum_j \frac{g_{j,31}(\alpha)}{j n_2 -(j-1) n_1}\bigg(\cos{\phi_j} \sin{\lambda_2} - \sin{\phi_j} \cos{\lambda_2 }\bigg)  \nonumber \\
 = 2\epsilon_1n_2 &\sum_j \frac{g_{j,31}(\alpha)}{j n_2 -(j-1) n_1}\sin{(\lambda_2-\phi_j)} \nonumber \\
 = 2\epsilon_1n_2&\sum_j \frac{g_{j,31}(\alpha) n_1 }{j n_2 -(j-1) n_1}\sin{\psi_{j-1}}\nonumber \\
 = 2 \epsilon_1n_2& \sum_k \frac{g_{k+1,31}(\alpha)}{(1+k)n_2-kn_1}\sin{\psi_k} = 2\epsilon_1 n_2 \sum_j \frac{g_{j+1,31}(\alpha)}{(1+j)n_2-jn_1}\sin{\psi_j} \nonumber \\
 = 2\epsilon_1 &\sum_j \frac{g_{j+1,31}(\alpha)}{1-\kappa}\sin{\psi_j}
\end{align}
Again, only the synodic angle $\psi_j$ remains! Then the expression for $\delta t_2$ is:
\begin{align}
\delta t_2 = -\frac{P_2}{2\pi}\epsilon_1\sum_{j=-\infty}^{j=\infty}& \bigg( \frac{d g_{j,0}(\alpha)}{d \alpha} \frac{2 \alpha }{\kappa}  +g_{j,0}(\alpha) \frac{3 j}{\kappa^2} +2 \frac{g_{j,0}(\alpha)}{\kappa}  +2  \frac{g_{j+1,31}(\alpha)}{1-\kappa}\bigg)\sin{\psi_j} \nonumber \\
 = -\frac{P_2}{2\pi}\epsilon_1\sum_{j=1}^{j=\infty}&\bigg( \bigg[\frac{d g_{j,0}(\alpha)}{d \alpha}+\frac{d g_{-j,0}(\alpha)}{d \alpha}\bigg]  \frac{2 \alpha }{\kappa}  +\bigg[g_{j,0}(\alpha)+g_{-j,0}(\alpha)\bigg] \bigg(\frac{3 j}{\kappa^2}+\frac{2}{\kappa}\bigg) \nonumber \\
&+2 \bigg[\frac{g_{j+1,31}(\alpha)}{1-\kappa}-\frac{g_{-j+1,31}(\alpha)}{1+\kappa}\bigg]\bigg)\sin{\psi_j} \nonumber \\
 = -\frac{P_2}{2\pi}\epsilon_1\sum_{j=1}^{j=\infty}&\bigg( \bigg[D_\alpha b^j_{1/2}(\alpha) - \alpha \delta_{j,1}\bigg]   \frac{2  }{\kappa}  +\bigg[ b^j_{1/2}(\alpha) - \alpha \delta_{j,1}\bigg]  \bigg(\frac{3 j+2\kappa}{\kappa^2}\bigg) \nonumber \\
&+2 \frac{(2j+\kappa) b^j_{1/2}(\alpha)+\kappa D_\alpha b^j_{1/2}(\alpha)-2\alpha \delta_{j+1,2}(1+\kappa)}{1-\kappa^2}\bigg)\sin{\psi_j} \nonumber \\
 = \frac{P_2}{2\pi}\epsilon_1\sum_{j=1}^{j=\infty} &\sin{\psi_j}\frac{j(\kappa^2+3)b^j_{1/2}(\alpha)+2\kappa(b^j_{1/2}(\alpha)+D_\alpha b^j_{1/2}(\alpha))-\alpha \delta_{j,1}(\kappa^2+4\kappa+3)}{\kappa^2(\kappa^2-1)}
\end{align}

However, this expression must be modified, since this corresponds to the orbit of the outer planet in Jacobi coordinates, about the center of mass of the inner planet and star subsystem. The correction to this is given in \citet{Agol2005} as:
\begin{align}
\delta t_{2,helio} = \delta t_{2,jacobi} -\frac{P_2}{2\pi} \epsilon_1 \alpha \sin{\psi_1} +O(e_1)
\end{align}
We add this to the indirect terms $I$ proportional to $\delta_{j,1}$ and rearrange:
\begin{align}
I&= -\delta_{j,1} \frac{\alpha(\kappa^2+4\kappa+3)+\alpha(\kappa^2(\kappa^2-1))}{\kappa^2(\kappa^2-1)} \nonumber \\
&= -\delta_{j,1} \frac{\alpha(\kappa^4+4\kappa+3)}{\kappa^2(\kappa^2-1)} \nonumber \\
& = -\delta_{j,1} \alpha^{-2}{(6+\alpha^{-3}-4\alpha^{-3/2})}{\kappa^2(\kappa^2-1)} \nonumber \\
& = -\delta_{j,1} \alpha^{-2}\frac{(\kappa^2-2\kappa+3)}{\kappa^2(\kappa^2-1)} 
\end{align}

so that we can write
\begin{align}\label{f2}
\delta t_2 & = \frac{P_2}{2\pi} \epsilon_1 \sum_{j=1}^{\infty} f_2^j(\alpha)\sin{\psi_j} \nonumber \\
f_2^j(\alpha) & = \frac{j(\kappa^2+3)b^j_{1/2}(\alpha)+2\kappa(b^j_{1/2}(\alpha)+D_\alpha b^j_{1/2}(\alpha))-\alpha^{-2} \delta_{j,1}(\kappa^2-2\kappa+3)}{\kappa^2(\kappa^2-1)}
\end{align}

We return now to one of the principle assumptions made in deriving these formulae, namely that all periodic terms in the Hamiltonian are short-period. Near and in the $j_R:j_R-1$ resonance this assumption breaks down. Indeed, the generating function itself does not converge because certain terms possess small denominators (the resonant combination of orbital frequencies). However, if we exclude those terms in the generating function, and proceed as we did in the derivation, the resulting TTV expression will still encompass the short-period effect. Moreover, they should be identical to our expressions but without the synodic terms with $j=j_R$ for the inner planet and $j=j_R-1$ for the outer planet. In other words, the synodic TTV expressions we derived should apply equally well for resonant and non-resonant systems with the {\it exception} that for systems near and in resonance the formula for the synodic harmonic aliased with the resonant frequency will be incorrect.

Finally, we also note that our resulting TTV expressions depend on the semimajor axis ratio and orbital periods of the {\it averaged} orbits. If the system is in the near resonant case, the timescale over which we average is longer. Therefore, if one observed a system for only a fraction of the super-period, the inferred average semimajor axis ratio could differ slightly from the ``true'' average one. This can lead to errors since the functions $f^{(j)}_i(\alpha)$ can be steep functions of $\alpha$ near the first order mean motion resonances.

\section{Convergence of synodic TTV sums}
Here we address the question of the convergence of the sums that appear in the synodic TTV expressions. We assume that the system is not near a resonance so that there are no small denominators and consider the limit of large $j$. The Laplace coefficients can be written as \citep{MurrayDermott}:
\begin{align}
\frac{1}{2}b^j_s(\alpha) &  = \frac{s(s+1)\ldots(s+j-1)}{j!}\alpha^j\bigg[ 1+\frac{s(s+j)}{1(j+1)}\alpha^2 +\frac{s(s+1)(s+j)(s+j+1)}{1\cdot2(j+1)(j+2)}\alpha^4+O(\alpha^6)\bigg]
\end{align}
In the limit of large $j$ and $\alpha<1$, this has a leading coefficient which scales as:
\begin{align}
b^j_s(\alpha) &  \propto \alpha^j
\end{align}
and therefore
\begin{align}
D_\alpha b^j_s(\alpha) &  \propto j\alpha^j
\end{align}
In the expression for $f_1^j$, given in Equation \ref{f1}, the quantities $\beta$ and $\beta^1-1$ appear, which in the limit of large $j$ have leading contributions of $\beta \propto j$ and $\beta^2 -1 \propto j^2$. In this limit, 
\begin{align}
f_1^j(\alpha) & = \alpha \frac{2\beta D_\alpha  b^j_{1/2}(\alpha) + j(3+\beta^2)  b^j_{1/2}(\alpha)-\alpha \delta_{j,1}(\beta^2+2\beta+3) }{\beta^2 (1-\beta^2)} \nonumber \\
f_1^j(\alpha)&\rightarrow \alpha \frac{2j^2\alpha^j+j(3+j^2)\alpha^j}{j^4}\nonumber \\
f_1^j(\alpha) &\rightarrow \alpha \frac{j^3\alpha^j}{j^4} = \frac{\alpha^{j+1}}{j}
\end{align} 

Similarly, for the outer planet, $\kappa \rightarrow j\alpha^{-3/2}$ and $1-\kappa^2 \rightarrow j^2 \alpha^{-3}$ so that
\begin{align}
f_2^j(\alpha) & = -\frac{j(\kappa^2+3)b^j_{1/2}(\alpha)+2\kappa(b^j_{1/2}(\alpha)+D_\alpha b^j_{1/2}(\alpha))-\alpha \delta_{j,1}(\kappa^2+4\kappa+3)}{\kappa^2(1-\kappa^2)} \nonumber \\
f_2^j(\alpha)&\rightarrow  \frac{j\kappa^2\alpha^j}{\kappa^4} \rightarrow  \frac{\alpha^{j+3}}{j}
\end{align} 

Therefore systems with smaller values of $\alpha$ will have smaller synodic TTVs and, given $\alpha$, the terms with larger $j$ are less important in determining the TTVs.

\end{document}